\documentclass[twocolumn]{aastex631}
\usepackage{amsmath}
\usepackage{epsf}
\usepackage{color}
\usepackage{graphicx}
\usepackage{color}
\usepackage{enumitem}
\usepackage{mathtools}
\usepackage[mathcal]{eucal}
 \let\mathscr\relax
\usepackage[scr]{rsfso}

\shorttitle{First Glimpse at Cosmic Dawn with {\it JWST} CEERS}
\shortauthors{Finkelstein et al.}

\newcommand{\sol}{$_{\odot}$}

\def\arcs{\hbox{$^{\prime\prime}$}}

\begin{document}
\title{A Long Time Ago in a Galaxy Far, Far Away: \\
A Candidate {\boldmath $z \sim$} 12 Galaxy in Early {\em JWST} CEERS Imaging}

\suppressAffiliations

\author[0000-0001-8519-1130]{Steven L. Finkelstein}
\affiliation{Department of Astronomy, The University of Texas at Austin, Austin, TX, USA}
\email{stevenf@astro.as.utexas.edu}

\author[0000-0002-9921-9218]{Micaela B. Bagley}
\affiliation{Department of Astronomy, The University of Texas at Austin, Austin, TX, USA}

%%%%%%%%%%%%%%%%%%%%%%%%%%%%%%%%%%%%%%%%%%%%%%%%%%%%%%%%%%%%%%%%%%%%%%%%%%%%
%%% Authors who provided significant contibutions over many days, beyond making figures %%%

\author[0000-0002-7959-8783]{Pablo Arrabal Haro}
\affiliation{NSF's National Optical-Infrared Astronomy Research Laboratory, 950 N. Cherry Ave., Tucson, AZ 85719, USA}

\author[0000-0001-5414-5131]{Mark Dickinson}
\affiliation{NSF's National Optical-Infrared Astronomy Research Laboratory, 950 N. Cherry Ave., Tucson, AZ 85719, USA}

\author[0000-0001-7113-2738]{Henry C. Ferguson}
\affiliation{Space Telescope Science Institute, 3700 San Martin Drive, Baltimore, MD 21218, USA}

\author[0000-0001-9187-3605]{Jeyhan S. Kartaltepe}
\affiliation{Laboratory for Multiwavelength Astrophysics, School of Physics and Astronomy, Rochester Institute of Technology, 84 Lomb Memorial Drive, Rochester, NY 14623, USA}

\author[0000-0001-7503-8482]{Casey Papovich}
\affiliation{Department of Physics and Astronomy, Texas A\&M University, College Station, TX, 77843-4242 USA}
\affiliation{George P.\ and Cynthia Woods Mitchell Institute for Fundamental Physics and Astronomy, Texas A\&M University, College Station, TX, 77843-4242 USA}

%%%%%%%%%%%%%%%%%%%%%%%%%%%%%%%%%%%%%%%%%%%%%%%%%%%%%%%%%%%%%%%%%%%%%%%%%%%%
%%% Authors who reduced/analyzed data / made figs / contributed data / etc
\author[0000-0002-4193-2539]{Denis Burgarella}
\affiliation{Aix Marseille Univ, CNRS, CNES, LAM Marseille, France}

\author[0000-0002-8360-3880]{Dale D. Kocevski}
\affiliation{Department of Physics and Astronomy, Colby College, Waterville, ME 04901, USA}

\author[0000-0002-1416-8483]{Marc Huertas-Company}
\affil{Instituto de Astrof\'isica de Canarias, La Laguna, Tenerife, Spain}
\affil{Universidad de la Laguna, La Laguna, Tenerife, Spain}
\affil{Universit\'e Paris-Cit\'e, LERMA - Observatoire de Paris, PSL, Paris, France}

\author[0000-0001-9298-3523]{Kartheik G. Iyer}
\affiliation{Dunlap Institute for Astronomy \& Astrophysics, University of Toronto, Toronto, ON M5S 3H4, Canada}

\author[0000-0002-6610-2048]{Anton M. Koekemoer}
\affiliation{Space Telescope Science Institute, 3700 San Martin Drive, Baltimore, MD 21218, USA}

\author[0000-0003-2366-8858]{Rebecca L. Larson}
\affiliation{NSF Graduate Fellow}
\affiliation{Department of Astronomy, The University of Texas at Austin, Austin, TX, USA}

\author[0000-0003-4528-5639]{Pablo G. P\'erez-Gonz\'alez}
\affiliation{Centro de Astrobiolog\'{\i}a (CAB/CSIC-INTA), Ctra. de Ajalvir km 4, Torrej\'on de Ardoz, E-28850, Madrid, Spain}

\author[0000-0002-8018-3219]{Caitlin Rose}
\affil{Laboratory for Multiwavelength Astrophysics, School of Physics and Astronomy, Rochester Institute of Technology, 84 Lomb Memorial Drive, Rochester, NY 14623, USA}

\author[0000-0002-8224-4505]{Sandro Tacchella}
\affiliation{Kavli Institute for Cosmology, University of Cambridge, Madingley Road, Cambridge, CB3 0HA, UK}\affiliation{Cavendish Laboratory, University of Cambridge, 19 JJ Thomson Avenue, Cambridge, CB3 0HE, UK}

\author[0000-0003-3903-6935]{Stephen M.~Wilkins} %
\affiliation{Astronomy Centre, University of Sussex, Falmer, Brighton BN1 9QH, UK}
\affiliation{Institute of Space Sciences and Astronomy, University of Malta, Msida MSD 2080, Malta}

%%%%%%%%%%%%%%%%%%%%%%%%%%%%%%%%%%%%%%%%%%%%%%%%%%%%%%%%%%%%%%%%%%%%%%%%%%%%%%
%%% CEERS members who worked on data prep crucial to our ability to make quick catalogs %%%
\author[0000-0003-4922-0613]{Katherine Chworowsky}\altaffiliation{NSF Graduate Fellow}
\affiliation{Department of Astronomy, The University of Texas at Austin, Austin, TX, USA}

\author{Aubrey Medrano}
\affiliation{Department of Astronomy, The University of Texas at Austin, Austin, TX, USA}

\author[0000-0003-4965-0402]{Alexa M.\ Morales}
\affiliation{Department of Astronomy, The University of Texas at Austin, Austin, TX, USA}

\author[0000-0002-6748-6821]{Rachel S.~Somerville}
\affiliation{Center for Computational Astrophysics, Flatiron Institute, 162 5th Avenue, New York, NY 10010, USA}

\author[0000-0003-3466-035X]{L. Y. Aaron\ Yung}
\affiliation{Astrophysics Science Division, NASA Goddard Space Flight Center, 8800 Greenbelt Rd, Greenbelt, MD 20771, USA}
%%%%%%%%%%%%%%%%%%%%%%%%%%%%%%%%%%%%%%%%%%%%%%%%%%%%%%%%%%%%%%%%%%%%%%%%%%%%%%
%%% CEERS architects and authors who contributed to the paper in other ways

\author[0000-0003-3820-2823]{Adriano Fontana}
\affiliation{INAF - Osservatorio Astronomico di Roma, via di Frascati 33, 00078 Monte Porzio Catone, Italy}

\author[0000-0002-7831-8751]{Mauro Giavalisco}
\affiliation{University of Massachusetts Amherst, 710 North Pleasant Street, Amherst, MA 01003-9305, USA}

\author[0000-0002-5688-0663]{Andrea Grazian}
\affiliation{INAF--Osservatorio Astronomico di Padova, Vicolo dell'Osservatorio 5, I-35122, Padova, Italy}

\author[0000-0001-9440-8872]{Norman A. Grogin}
\affiliation{Space Telescope Science Institute, 3700 San Martin Drive, Baltimore, MD 21218, USA}

\author[0000-0001-8152-3943]{Lisa J. Kewley}
\affiliation{Harvard-Smithsonian Center for Astrophysics, 60 Garden Street, Cambridge, MA 02138, USA}

\author[0000-0002-5537-8110]{Allison Kirkpatrick}
\affiliation{Department of Physics and Astronomy, University of Kansas, Lawrence, KS 66045, USA}

\author[0000-0002-8816-5146]{Peter Kurczynski}
\affiliation{Observational Cosmology Laboratory, Code 665, NASA Goddard Space Flight Center, Greenbelt, MD 20771}
 
\author[0000-0003-3130-5643]{Jennifer M. Lotz}
\affiliation{Gemini Observatory/NSF's National Optical-Infrared Astronomy Research Laboratory, 950 N. Cherry Ave., Tucson, AZ 85719, USA}

\author[0000-0001-8940-6768]{Laura Pentericci}
\affiliation{INAF - Osservatorio Astronomico di Roma, via di Frascati 33, 00078 Monte Porzio Catone, Italy}

\author[0000-0003-3382-5941]{Nor Pirzkal}
\affiliation{ESA/AURA Space Telescope Science Institute}

\author[0000-0002-5269-6527]{Swara Ravindranath}
\affiliation{Space Telescope Science Institute, 3700 San Martin Drive, Baltimore, MD 21218, USA}

\author[0000-0003-0894-1588]{Russell E.\ Ryan Jr.}
\affiliation{Space Telescope Science Institute, 3700 San Martin Drive, Baltimore, MD 21218, USA}

\author[0000-0002-1410-0470]{Jonathan R. Trump}
\affiliation{Department of Physics, 196 Auditorium Road, Unit 3046, University of Connecticut, Storrs, CT 06269, USA}

\author[0000-0001-8835-7722]{Guang Yang}
\affiliation{Kapteyn Astronomical Institute, University of Groningen, P.O. Box 800, 9700 AV Groningen, The Netherlands}
\affiliation{SRON Netherlands Institute for Space Research, Postbus 800, 9700 AV Groningen, The Netherlands}

%%%%%%%%%%%%%%%%%%%%%%%%%%%%%%%%%%%%%%%%%%%%%%%%%%%%%%%%%%%%%%%%%%%%%%%%%%%%%%
%%% CEERS team co-authors here (alphabetically)
\collaboration{85}{and The CEERS Team:}
\author[0000-0001-9328-3991]{Omar Almaini}
\affiliation{School of Physics and Astronomy, University of Nottingham, University Park, Nottingham NG7 2RD, UK}

\author[0000-0001-5758-1000]{Ricardo O. Amor\'{i}n}
\affiliation{Instituto de Investigaci\'{o}n Multidisciplinar en Ciencia y Tecnolog\'{i}a, Universidad de La Serena, Raul Bitr\'{a}n 1305, La Serena 2204000, Chile}
\affiliation{Departamento de Astronom\'{i}a, Universidad de La Serena, Av. Juan Cisternas 1200 Norte, La Serena 1720236, Chile}

\author[0000-0002-8053-8040]{Marianna Annunziatella}
\affiliation{Centro de Astrobiolog\'ia (CSIC-INTA), Ctra de Ajalvir km 4, Torrej\'on de Ardoz, 28850, Madrid, Spain}

\author[0000-0001-8534-7502]{Bren E. Backhaus}
\affiliation{Department of Physics, 196 Auditorium Road, Unit 3046, University of Connecticut, Storrs, CT 06269}

\author[0000-0002-0786-7307]{Guillermo Barro}
\affiliation{Department of Physics, University of the Pacific, Stockton, CA 90340 USA}

\author[0000-0002-2517-6446]{Peter Behroozi}
\affiliation{Department of Astronomy and Steward Observatory, University of Arizona, Tucson, AZ 85721, USA}
\affiliation{Division of Science, National Astronomical Observatory of Japan, 2-21-1 Osawa, Mitaka, Tokyo 181-8588, Japan}

\author[0000-0002-5564-9873]{Eric F.\ Bell}
\affiliation{Department of Astronomy, University of Michigan, 1085 S. University Ave, Ann Arbor, MI 48109-1107, USA}

\author[0000-0003-0883-2226]{Rachana Bhatawdekar}
\affiliation{European Space Agency, ESA/ESTEC, Keplerlaan 1, 2201 AZ Noordwijk, NL}

\author[0000-0003-0492-4924]{Laura Bisigello}
\affiliation{Dipartimento di Fisica e Astronomia "G.Galilei", Universit\'a di Padova, Via Marzolo 8, I-35131 Padova, Italy}
\affiliation{INAF--Osservatorio Astronomico di Padova, Vicolo dell'Osservatorio 5, I-35122, Padova, Italy}

\author[0000-0003-0212-2979]{Volker Bromm}
\affiliation{Department of Astronomy, The University of Texas at Austin, Austin, TX, USA}

\author[0000-0003-3441-903X]{V\'eronique Buat}
\affiliation{Aix Marseille Univ, CNRS, CNES, LAM Marseille, France}

\author[0000-0002-2861-9812]{Fernando Buitrago}
\affiliation{Departamento de F\'{i}sica Te\'{o}rica, At\'{o}mica y \'{O}ptica, Universidad de Valladolid, 47011 Valladolid, Spain}
\affiliation{Instituto de Astrof\'{i}sica e Ci\^{e}ncias do Espa\c{c}o, Universidade de Lisboa, OAL, Tapada da Ajuda, PT1349-018 Lisbon, Portugal}

\author[0000-0003-2536-1614]{Antonello Calabr{\`o}}
\affiliation{Osservatorio Astronomico di Roma, via Frascati 33, Monte Porzio Catone, Italy}

\author[0000-0002-0930-6466]{Caitlin M.\ Casey}
\affiliation{Department of Astronomy, The University of Texas at Austin, Austin, TX, USA}

\author[0000-0001-9875-8263]{Marco Castellano}
\affiliation{INAF - Osservatorio Astronomico di Roma, via di Frascati 33, 00078 Monte Porzio Catone, Italy}

\author[0000-0003-2332-5505]{\'Oscar A. Ch\'avez Ortiz}
\affiliation{Department of Astronomy, The University of Texas at Austin, Austin, TX, USA}

\author[0000-0003-0541-2891]{Laure Ciesla}
\affiliation{Aix Marseille Univ, CNRS, CNES, LAM Marseille, France}

\author[0000-0001-7151-009X]{Nikko J. Cleri}
\affiliation{Department of Physics and Astronomy, Texas A\&M University, College Station, TX, 77843-4242 USA}
\affiliation{George P.\ and Cynthia Woods Mitchell Institute for Fundamental Physics and Astronomy, Texas A\&M University, College Station, TX, 77843-4242 USA}

\author[0000-0003-3329-1337]{Seth H. Cohen}
\affiliation{School of Earth and Space Exploration, Arizona State University, Tempe, AZ, 85287 USA}

\author[0000-0002-6348-1900]{Justin W. Cole}
\affiliation{Department of Physics and Astronomy, Texas A\&M University, College Station, TX, 77843-4242 USA}
\affiliation{George P.\ and Cynthia Woods Mitchell Institute for Fundamental Physics and Astronomy, Texas A\&M University, College Station, TX, 77843-4242 USA}

\author[0000-0002-2200-9845]{Kevin C. Cooke}
\affiliation{AAAS S\&T Policy Fellow hosted at the National Science Foundation, 1200 New York Ave, NW, Washington, DC, US 20005}

\author[0000-0003-1371-6019]{M. C. Cooper}
\affiliation{Department of Physics \& Astronomy, University of California, Irvine, 4129 Reines Hall, Irvine, CA 92697, USA}

\author[0000-0002-3892-0190]{Asantha R. Cooray}
\affiliation{Department of Physics \& Astronomy, University of
California, Irvine, 4129 Reines Hall, Irvine, CA 92697, USA}

\author[0000-0001-6820-0015]{Luca Costantin}
\affiliation{Centro de Astrobiolog\'ia (CSIC-INTA), Ctra de Ajalvir km 4, Torrej\'on de Ardoz, 28850, Madrid, Spain}

\author[0000-0002-1803-794X]{Isabella G. Cox}
\affiliation{Laboratory for Multiwavelength Astrophysics, School of Physics and Astronomy, Rochester Institute of Technology, 84 Lomb Memorial Drive, Rochester, NY 14623, USA}

\author[0000-0002-5009-512X]{Darren Croton}
\affiliation{Centre for Astrophysics \& Supercomputing, Swinburne University of Technology, Hawthorn, VIC 3122, Australia}
\affiliation{ARC Centre of Excellence for All Sky Astrophysics in 3 Dimensions (ASTRO 3D)}

\author[0000-0002-3331-9590]{Emanuele Daddi}
\affiliation{Universit\'e Paris-Saclay, Universit\'e Paris Cit\'e, CEA, CNRS, AIM, 91191, Gif-sur-Yvette, France}

\author[0000-0003-2842-9434]{Romeel Dav\'e}
\affiliation{Institute for Astronomy, University of Edinburgh, Blackford Hill, Edinburgh, EH9 3HJ UK}
\affiliation{Department of Physics and Astronomy, University of the Western Cape, Robert Sobukwe Rd, Bellville, Cape Town 7535, South Africa}

\author[0000-0002-6219-5558]{Alexander de la Vega}
\affiliation{Department of Physics and Astronomy, Johns Hopkins University, Baltimore, MD, USA}

\author[0000-0003-4174-0374]{Avishai Dekel}
\affil{Racah Institute of Physics, The Hebrew University of Jerusalem, Jerusalem 91904, Israel}

\author[0000-0002-7631-647X]{David Elbaz}
\affil{Universit{\'e} Paris-Saclay, Université Paris Cit{\'e}, CEA, CNRS, AIM, 91191, Gif-sur-Yvette, France}

\author[0000-0001-8489-2349]{Vicente Estrada-Carpenter}
\affiliation{Department of Astronomy \& Physics, Saint Mary's University, 923 Robie Street, Halifax, NS, B3H 3C3, Canada}

\author{Sandra M.\ Faber}
\affiliation{University of California Observatories and Department of Astronomy and Astrophysics, University of California, Santa Cruz, 1156 High Street, Santa Cruz, CA 95064, USA}

\author[0000-0003-0531-5450]{Vital Fern\'{a}ndez}
\affiliation{Instituto de Investigaci\'{o}n Multidisciplinar en Ciencia y Tecnolog\'{i}a, Universidad de La Serena, Raul Bitr\'{a}n 1305, La Serena 2204000, Chile}

\author[0000-0003-0792-5877]{Keely D. Finkelstein}
\affiliation{Department of Astronomy, The University of Texas at Austin, Austin, TX, USA}

\author[0000-0002-5245-7796]{Jonathan Freundlich}
\affiliation{Université de Strasbourg, CNRS, Observatoire Astronomique de Strasbourg, UMR 7550, F-67000 Strasbourg, France}

\author[0000-0001-7201-5066]{Seiji Fujimoto}
\affiliation{Cosmic Dawn Center (DAWN), Jagtvej 128, DK2200 Copenhagen N, Denmark}
\affiliation{Niels Bohr Institute, University of Copenhagen, Lyngbyvej 2, DK2100 Copenhagen \O, Denmark}

\author[0000-0002-8365-5525]{\'Angela Garc\'ia-Argum\'anez}
\affiliation{Departamento de Física de la Tierra y Astrofísica, Facultad de CC Físicas, Universidad Complutense de Madrid, E-28040, Madrid, Spain}
\affiliation{Instituto de Física de Partículas y del Cosmos IPARCOS, Facultad de CC Físicas, Universidad Complutense de Madrid, 28040 Madrid, Spain}

\author[0000-0003-2098-9568]{Jonathan P. Gardner}
\affiliation{Astrophysics Science Division, NASA Goddard Space Flight Center, 8800 Greenbelt Rd, Greenbelt, MD 20771, USA}

\author[0000-0003-1530-8713]{Eric Gawiser}
\affiliation{Department of Physics and Astronomy, Rutgers, the State University of New Jersey, Piscataway, NJ 08854, USA}

\author[0000-0002-4085-9165]{Carlos G{\'o}mez-Guijarro}
\affil{Universit{\'e} Paris-Saclay, Universit{\'e} Paris Cit{\'e}, CEA, CNRS, AIM, 91191, Gif-sur-Yvette, France}

\author[0000-0002-4162-6523]{Yuchen Guo}
\affiliation{Department of Astronomy, The University of Texas at Austin, Austin, TX, USA}

\author[0000-0002-6292-4589]{Kurt Hamblin}
\affiliation{Department of Physics and Astronomy, University of Kansas, Lawrence, KS 66045, USA}

\author[0000-0002-9753-1769]{Timothy S. Hamilton}
\affiliation{Shawnee State University, Portsmouth, OH, USA}

\author[0000-0001-6145-5090]{Nimish P. Hathi}
\affiliation{Space Telescope Science Institute, 3700 San Martin Drive, Baltimore, MD 21218, USA}

\author[0000-0002-4884-6756]{Benne W. Holwerda}
\affil{Physics \& Astronomy Department, University of Louisville, 40292 KY, Louisville, USA}

\author[0000-0002-3301-3321]{Michaela Hirschmann}
\affiliation{Institute of Physics, Laboratory of Galaxy Evolution, Ecole Polytechnique Fédérale de Lausanne (EPFL), Observatoire de Sauverny, 1290 Versoix, Switzerland}

\author[0000-0001-6251-4988]{Taylor A. Hutchison}
\affiliation{NSF Graduate Fellow}
\affiliation{Department of Physics and Astronomy, Texas A\&M University, College Station, TX, 77843-4242 USA}
\affiliation{George P.\ and Cynthia Woods Mitchell Institute for Fundamental Physics and Astronomy, Texas A\&M University, College Station, TX, 77843-4242 USA}

\author[0000-0002-6790-5125]{Anne E. Jaskot}
\affiliation{Department of Astronomy, Williams College, Williamstown, MA, 01267, USA}

\author[0000-0001-8738-6011]{Saurabh W. Jha}
\affiliation{Department of Physics and Astronomy, Rutgers, the State University of New Jersey, Piscataway, NJ 08854, USA}

\author[0000-0002-1590-0568]{Shardha Jogee}
\affiliation{Department of Astronomy, The University of Texas at Austin, Austin, TX, USA}

\author[0000-0002-0000-2394]{St{\'e}phanie Juneau}
\affiliation{NSF's NOIRLab, 950 N. Cherry Ave., Tucson, AZ 85719, USA}

\author[0000-0003-1187-4240]{Intae Jung}
\affil{Department of Physics, The Catholic University of America, Washington, DC 20064, USA }
\affil{Astrophysics Science Division, NASA Goddard Space Flight Center, 8800 Greenbelt Rd, Greenbelt, MD 20771, USA}
\affil{Center for Research and Exploration in Space Science and Technology, NASA/GSFC, Greenbelt, MD 20771}

\author{Susan A. Kassin}
\affiliation{Space Telescope Science Institute, Baltimore, MD, 21218, USA}
\affiliation{Dept. of Physics \& Astronomy, Johns Hopkins University, 3400 N. Charles St., Baltimore, MD, 21218, USA}

\author[0000-0002-9466-2763]{Aur{\'e}lien Le Bail}
\affil{Universit{\'e} Paris-Saclay, Université Paris Cit{\'e}, CEA, CNRS, AIM, 91191, Gif-sur-Yvette, France}

\author[0000-0002-9393-6507]{Gene C. K. Leung}
\affiliation{Department of Astronomy, The University of Texas at Austin, Austin, TX, USA}

\author[0000-0003-1581-7825]{Ray A. Lucas}
\affiliation{Space Telescope Science Institute, 3700 San Martin Drive, Baltimore, MD 21218, USA}

\author[0000-0002-6777-6490]{Benjamin Magnelli}
\affiliation{Universit\'e Paris-Saclay, Universit\'e Paris Cit\'e, CEA, CNRS, AIM, 91191, Gif-sur-Yvette, France}

\author{Kameswara Bharadwaj Mantha}
\affiliation{Minnesota Institute for Astrophysics, University of Minnesota, 116 church St SE, Minneapolis, MN, 55455, USA.}

\author[0000-0002-7547-3385]{Jasleen Matharu}
\affiliation{Department of Physics and Astronomy, Texas A\&M University, College Station, TX, 77843-4242 USA}
\affiliation{George P.\ and Cynthia Woods Mitchell Institute for Fundamental Physics and Astronomy, Texas A\&M University, College Station, TX, 77843-4242 USA}

\author[0000-0001-8688-2443]{Elizabeth J.\ McGrath}
\affiliation{Department of Physics and Astronomy, Colby College, Waterville, ME 04901, USA}

\author{Daniel H. McIntosh}
\affiliation{Division of Energy, Matter and Systems, School of Science and Engineering, University of Missouri-Kansas City, Kansas City, MO 64110, USA}

\author[0000-0001-6870-8900]{Emiliano Merlin}
\affiliation{INAF Osservatorio Astronomico di Roma, Via Frascati 33, 00078 Monteporzio Catone, Rome, Italy}

\author{Bahram Mobasher}
\affiliation{Department of Physics and Astronomy, University of California, 900 University Ave, Riverside, CA 92521, USA}

\author[0000-0001-8684-2222]{Jeffrey A.\ Newman}
\affiliation{Department of Physics and Astronomy and PITT PACC, University of Pittsburgh, Pittsburgh, PA 15260, USA}

\author[0000-0003-0892-5203]{David C. Nicholls}
\affiliation{Research School of Astronomy and Astrophysics, Australian National University, Canberra, ACT 2600, Australia}

\author[0000-0002-2499-9205]{Viraj Pandya}
\altaffiliation{Hubble Fellow}
\affiliation{Columbia Astrophysics Laboratory, Columbia University, 550 West 120th Street, New York, NY 10027, USA}

\author[0000-0002-9946-4731]{Marc Rafelski}
\affiliation{Space Telescope Science Institute, 3700 San Martin Drive, Baltimore, MD 21218, USA}
\affiliation{Department of Physics and Astronomy, Johns Hopkins University, Baltimore, MD 21218, USA}

\author[0000-0001-5749-5452]{Kaila Ronayne}
\affiliation{Department of Physics and Astronomy, Texas A\&M University, College Station, TX, 77843-4242 USA}
\affiliation{George P.\ and Cynthia Woods Mitchell Institute for Fundamental Physics and Astronomy, Texas A\&M University, College Station, TX, 77843-4242 USA}

\author[0000-0002-9334-8705]{Paola Santini}
\affiliation{INAF - Osservatorio Astronomico di Roma, via di Frascati 33, 00078 Monte Porzio Catone, Italy}

\author[0000-0001-7755-4755]{Lise-Marie Seill\'e}
\affiliation{Aix Marseille Univ, CNRS, CNES, LAM Marseille, France}

\author[0000-0001-7811-9042]{Ekta A. Shah}
\affiliation{Department of Physics and Astronomy, University of California,Davis, One Shields Ave, Davis, CA 95616, USA}

\author[0000-0001-9495-7759]{Lu Shen}
\affil{CAS Key Laboratory for Research in Galaxies and Cosmology, Department of Astronomy, University of Science and Technology of China, Hefei 230026, China}
\affil{School of Astronomy and Space Sciences, University of Science and Technology of China, Hefei, 230026, China}

\author[0000-0002-6386-7299]{Raymond C. Simons}
\affiliation{Space Telescope Science Institute, 3700 San Martin Drive, Baltimore, MD 21218, USA}

\author{Gregory F. Snyder}
\affiliation{Space Telescope Science Institute, 3700 San Martin Drive, Baltimore, MD 21218, USA}

\author[0000-0002-8770-809X]{Elizabeth R. Stanway}
\affiliation{Department of Physics, University of Warwick, Coventry, CV4 7AL, United Kingdom}

\author[0000-0002-4772-7878]{Amber N. Straughn}
\affiliation{Astrophysics Science Division, NASA Goddard Space Flight Center, 8800 Greenbelt Rd, Greenbelt, MD 20771, USA}

\author[0000-0002-7064-5424]{Harry I. Teplitz}
\affiliation{IPAC, Mail Code 314-6, California Institute of Technology, 1200 E. California Blvd., Pasadena CA, 91125, USA}

\author[0000-0002-8163-0172]{Brittany N. Vanderhoof}
\affil{Laboratory for Multiwavelength Astrophysics, School of Physics and Astronomy, Rochester Institute of Technology, 84 Lomb Memorial Drive, Rochester, NY 14623, USA}

\author[0000-0003-2338-5567]{Jes\'us Vega-Ferrero}
\affil{Instituto de Astrof\'isica de Canarias, La Laguna, Tenerife, Spain}

\author[0000-0002-9593-8274]{Weichen Wang}
\affiliation{Department of Physics and Astronomy, Johns Hopkins University, 3400 N. Charles Street, Baltimore, MD 21218, USA}

\author[0000-0001-6065-7483]{Benjamin J. Weiner}
\affiliation{MMT/Steward Observatory, University of Arizona, 933 N. Cherry St, Tucson, AZ 85721, USA}

\author[0000-0001-9262-9997]{Christopher N. A. Willmer}
\affiliation{Steward Observatory, University of Arizona, 933 N.\ Cherry Ave, Tucson, AZ 85721,USA}

\author[0000-0003-3735-1931]{Stijn Wuyts}
\affiliation{Department of Physics, University of Bath, Claverton Down, Bath BA2 7AY, UK}

\author[0000-0002-0786-7307]{Jorge A. Zavala}
\affiliation{National Astronomical Observatory of Japan, 2-21-1 Osawa, Mitaka, Tokyo 181-8588, Japan}

%%% end authors

\begin{abstract} 
We report the discovery of a candidate galaxy with a photo-z of $z
\sim$ 12 in the first epoch of the {\it JWST} Cosmic Evolution Early
Release Science (CEERS) Survey. Following conservative selection
criteria we identify a source with a robust $z_{phot} =$
  11.8$^{+0.3}_{-0.2}$ (1$\sigma$ uncertainty) with m$_{F200W}
  =$ 27.3, and $\gtrsim$7$\sigma$ detections in five filters. 
The source is not detected at $\lambda <$ 1.4$\mu$m in deep imaging
from both {\it HST} and {\it JWST}, and has faint $\sim$3$\sigma$
detections in {\it JWST} F150W and {\it HST} F160W, which signal a
Ly$\alpha$ break near the red edge of both filters, implying $z \sim$ 12.
This object (Maisie's Galaxy) exhibits F115W$-$F200W $>$ 1.9 mag
  (2$\sigma$ lower limit) with a blue continuum slope, resulting in
99.6\% of the photo-z PDF favoring $z >$ 11. All data quality
images show no artifacts at the candidate's position, and independent
analyses consistently find a strong preference for $z >$ 11. 
Its colors are inconsistent with Galactic stars, and it is resolved
($r_{h} =$ 340 $+/-$ 14 pc). Maisie's Galaxy has log
M$_{\ast}$/M\sol $\sim$ 8.5 and is highly star-forming (log
sSFR$\sim$$-$8.2 yr$^{-1}$), with a blue rest-UV color ($\beta \sim
-$2.5) indicating little dust though not extremely low
metallicity. While the presence of this source is in tension with
most predictions, it agrees with empirical extrapolations assuming UV
luminosity functions which smoothly decline with increasing redshift. Should followup spectroscopy validate
this redshift, our Universe was already aglow with galaxies less than 400 Myr after the Big Bang.
 \end{abstract}


\section{Introduction}\label{sec:intro}

The study of galaxy evolution is the ultimate human origin story -- not just how did our species, planet or Solar System come to be, but this field seeks to answer how our Milky Way Galaxy came to be.  One method to study our Galactic origins is to study the earliest building blocks of the Milky Way by searching for and analyzing galaxies forming in the early Universe.  The advent of the Wide Field Camera 3 (WFC3) on the {\it Hubble Space Telescope} ({\it HST}) pushed our cosmic horizons well into the epoch of reionization, the time when energetic photons (presumably from massive stars in early galaxies) ionized the gas in the intergalactic medium \citep[IGM; e.g.][and references therein]{finkelstein16, stark16, robertson21}.  These studies found that the $z =$ 6--10 universe is teeming with galaxies, with thousands of galaxy candidates known, including spectroscopic confirmations out to $z \sim$ 11 \citep{oesch16,jiang21}.

One key focus in these studies has been the evolution of the cosmic star-formation rate density (SFRD).  This quantity is well known to rise from the present day to the peak of cosmic star-formation at $z \sim$ 2--3, then decline again to early times \citep[e.g.][]{madau14}.  As the aforementioned WFC3 studies pushed to higher redshifts, it became of interest to study whether the cosmic SFRD, which exhibited a smooth decline from $z =$ 4--8 \citep[e.g.][]{bouwens15,finkelstein15}, continued to decline smoothly to even higher redshifts.  Results in the literature were mixed, with some studies finding evidence for an accelerated decline in the SFRD \citep[e.g.][]{oesch18,bouwens21}, while others found that observations supported a continued smooth decline \citep[e.g.][]{coe13,mcleod16,finkelstein22b}.
Simulations do make predictions for the evolution of the SFRD, but these predictions span a wide range \citep[e.g.][]{gnedin16,dayal18,tacchella18,yung19a,behroozi20}.

Part of the difficulty of such studies is the near-heroic observational effort needed to study galaxies at $z \sim$ 10 with {\it HST}.  These galaxies become more and more difficult to see with this 2.4m ultraviolet (UV)/optical/near-IR telescope, and at these high-redshifts they become single-band detections, leaving the $z \gtrsim$ 11 universe opaque to our understanding.  To avoid being dominated by spurious sources, studies employ a variety of vetting criteria to ensure robust samples of candidate galaxies \citep[e.g.][]{bouwens21,bagley22,finkelstein22}, which makes it difficult to estimate the sample completeness and thereby to obtain a robust estimate of the SFRD. 

This all changes with the advent of the {\it James Webb Space Telescope} ({\it JWST}).  The dramatic increase in light-gathering power coupled with the infrared sensitivity makes this telescope the ideal machine to push our cosmic horizons to the epoch of the first galaxies.  As the first {\it JWST} images arrive it is natural to wonder what these early data tell us about the rise of star-formation in the early universe.  If the SFRD really declines as steeply at $z >$ 8 as has been proposed, few galaxies at $z >$ 11 should be detectable in early {\it JWST} data.  If the decline is instead more gradual one might expect to discover galaxies at $z \sim$ 12 or even higher.  In just the first week since the data have been released exciting results already indicate significant star formation is occurring at $z >$ 11 \citep[e.g.][]{castellano22,naidu22}. 

As another early probe of this epoch, here we report on a search for the highest redshift ($z \gtrsim$ 12) galaxies in the first epoch of imaging from the Cosmic Evolution Early Release Science Survey (CEERS; Finkelstein et al.\ in prep).  These data were among the first Cycle 1 science exposures taken, and were included in the first publicly released data on July 14.  \S 2 describes the observations and data reduction, while \S 3 describes our photometry, photometric redshift measurement, and sample selection procedure. \S 4 presents our results, and we discuss these results in \S 5.  Our conclusions are presented in \S 6.  In this paper we assume the latest {\it Planck} flat $\Lambda$CDM cosmology with H$_{0}=$67.36, $\Omega_m=$0.3153, and $\Omega_{\Lambda}=$0.6847 \citep{planck20}.  All magnitudes are in the absolute bolometric system \citep[AB][]{oke83}.

\section{Observations}

\subsection{CEERS Data}

CEERS is one of 13 early release science surveys designed to obtain
data covering all areas of astronomy early in Cycle 1.  CEERS is based
around a mosaic of 10 NIRCam \citep{rieke05} pointings, with six
obtaining NIRSpec \citep{nirspec} in parallel, and four with MIRI
\citep{miri} in parallel (four of these pointings also include NIRCam
wide-field slitless grism spectroscopy; \citealt{nircamgrism}).  Here
we make use of the first four CEERS NIRCam pointings, obtained on 21
June 2022, known as CEERS1, CEERS2, CEERS3, and CEERS6.

In each pointing, data were obtained in the short-wavelength (SW) channel F115W, F150W, and F200W filters, and long-wavelength (LW) channel F277W, F356W, F410M, and F444W filters.  The total exposure time for pixels observed in all three dithers was typically 2835 s per filter.  The exception is F115W, which obtained double the exposure time to increase the depth on the filter covering the wavelength range below the Lyman-$\alpha$ break at $z >$ 10.  The full details on the readout and dither patterns will be available in the CEERS overview paper (Finkelstein et al.\ in prep). 

\subsection{Data Reduction}
We performed a careful initial reduction of the NIRCam images in all
four pointings, using version 1.6.2 of the \textit{JWST} Calibration
Pipeline\footnote{\url{jwst-pipeline.readthedocs.io}} with some custom
modifications. We used the current (29 July 2022) set of NIRCam
reference files\footnote{\url{jwst-crds.stsci.edu}, jwst\_nircam\_0221.imap},
which includes in flight readnoise, superbias, distortion, and 
photometric flux calibration references. We note that the flats were created
pre-flight. We describe our reduction steps below,
and present more details in Bagley et al. (in prep).

Beginning with the raw data, we used Stage 1 of the pipeline with all default
parameters to apply detector-level corrections, fit the ramps in each
integration, and output countrate maps. We next subtracted the ``wisp''
features, stray light that is reflected off the secondary mirror supports,
from detectors A3, B3 and B4 for filters F150W and F200W. For each image $I$,
we scaled the corresponding wisp template\footnote{\url{jwst-docs.stsci.edu/jwst-near-infrared-camera/nircam-features-and-caveats/nircam-claws-and-wisps}} $W$
(available as of 8 July 2022) by the coefficient $a$ that minimized
Var$(I - a W)$, and subtracted the scaled template.
We then performed a custom step to remove \textit{1/f} noise, which is
correlated noise introduced in the images during the detector readout that
presents as horizontal and vertical striping patterns \citep{schlawin20}.
We applied the flat field to the countrate maps to ensure we were measuring
the \textit{1/f} noise pattern on a flat image. We masked all bad pixels
and source flux, using \textsc{Photutils} \citep{photutils} to detect sources
and implementing a tiered approach to source masking. This approach 
convolves the image with progressively smaller kernels, identifying sources
at each step. We use four tiers, with Gaussian kernels of $\sigma=25$, 15, 
5 and 2 pixels (on the original 0\farcs031/pixel and 0\farcs063/pixel scales
for the SW and LW channels, respectively). These values were chosen after 
experimenting with several filter kernels to aggressively mask as much source
flux as possible. First for each row and then each column, we measured a
sigma-clipped median value and subtracted this value from the un-flat-fielded
countrate map. This correction was performed amplifier-by-amplifier in
all filters except F444W, for which we did not perform any correction because
the significant residual flat field structure present in the images dominated
any $1/f$ pattern.

After processing the cleaned countrate maps through Stage 2 of the pipeline,
we performed an astrometric calibration using an edited version of the
TweakReg step of the pipeline. The TweakReg step detects sources in each
input image, identifies their counterparts in the reference catalog, and
calculates a transformation to correct the image WCS.
We ran \textsc{Source Extractor} \citep{bertin96} on each individual
image to replace the internal TweakReg source identification, finding that
\textsc{Source Extractor} did a better job of identifying and deblending
real sources.
In lieu of using the default options that allow for alignment to Gaia DR2,
we used a reference catalog derived from a \textit{HST} F160W 0\farcs03/pixel
mosaic\footnote{\url{ceers.github.io/hdr1.html}} in the EGS field with
astrometry tied to Gaia-EDR3 \citep[see][for details]{koekemoer11}.
We first determined relative offsets between images of the same
detector, allowing for shifts in $x$ and $y$. The RMS of this relative
astrometry is $\sim$3-6 mas. We then aligned all images to our
\textit{HST} F160W reference catalog, allowing for shifts in $x$ and $y$,
rotations and, in the LW images only, a scaling to account for any additional
distortion (though we note that the scaling factor is $\sim1\times10^{-5}$).
The RMS of this absolute alignment is $\sim12-15$mas, and the
alignment between NIRCam images in different filters has an RMS of $5-10$mas.

We fit and removed a single value in MJy/sr from each calibrated detector image
separately before coadding the images onto a common output grid.
For each image, we additionally calculate a scaling factor for the
readnoise variance array (\texttt{VAR\_RNOISE}) from the background-subtracted
sky pixels, again avoiding source flux and bad pixels. We apply this scaling
factor to the variance arrays so that they include the robustly-measured
sky variance. The coadding was performed using the drizzle algorithm with an inverse variance
map weighting \citep{fruchter02,casertano00} via the Resample step in the
pipeline. The output mosaics have pixels scales of 0.03\arcs /pixel. The usable total area covered by these observations,
calculated from the number of pixels with low effective error-map
values in all of the F115W, F150W, F200W, F277W, and the detection
image (see below) is 34.5 arcmin$^2$.

We note that our data reduction represents a preliminary version, with several
aspects that will be improved with the release of updated NIRCam reference
files. We also have not removed the features known as ``snowballs'' from the mosaics at this time. However, we have carefully inspected all input exposures to
ensure that the fluxes in all filters at the positions of galaxies of interest are unaffected by snowballs (see
Section~\ref{sec:fidelity}).

\section{Methodology}

\subsection{Photometric Catalog Construction}\label{sec:phot}

The full details of our photometric analysis will be presented in
Finkelstein et al.\ (2022d, in prep); here we briefly
summarize our procedures (many of which are similar to
\citealt{finkelstein22}).  The data products from our modified data
reduction pipeline come in the form of multi-extension ``i2d" files.
We first estimate and subtract any residual background using a custom
Python-based algorithm.  This routine iteratively convolves the image
with Gaussian kernels of progressively smaller sizes (with $\sigma=$25, 15, 5
  and 2 pixels), then uses
\textsc{Photutils} to mask pixels identified with sources in four
iterations to mask progressively smaller sources, dilating the masks
in between iterations (by 33, 25, 21 and 19 pixels), then measuring the background after masking
with \textsc{photutils.Background2D}.  This final background
  image-construction step used the BkgZoomInterpolator algorithm to
  construct a smooth background based on the robust sigma-clipped
  means measured within boxes of 10x10 pixels, and median-filtered over
  5x5 adjacent boxes.

The i2d file was split into separate extensions, subtracting this background from the SCI extension.  Empirical PSFs were made by stacking stars, and the F115W, F150W, F200W, and F277W images were PSF-matched to the F356W image using \textsc{pypher}.  Photometry was computed on the PSF-matched images using \textsc{Source Extractor} \citep[hereafter SE;][]{bertin96} v2.25.0 in two image mode, with an inverse-variance weighted combination of the PSF-matched F277W and F356W images as the detection image, with photometry measured on all seven bands.

Colors were measured in small Kron apertures with a Kron factor of 0.8
and a Kron minimum radius of 1.1 pixels; this is smaller than previous studies, which we found necessary to keep the elliptical aperture close to the significant isophotes of small, faint galaxies.  An aperture correction was derived in the F356W catalog as the ratio between the flux measured in the default Kron aperture (with PHOT\_AUTOPARAMS 2.5, 3.5) to that in our small Kron aperture.  This correction was applied to all fluxes and uncertainties.  We use the CEERS simulated imaging\footnote{Simulated Data Release 3; \url{ceers.github.io/sdr3.html}} to test the accuracy of this procedure, finding that after this aperture correction, total fluxes were underestimated by $\sim$10-15\%, rising to 22\% in F444W (understandable due to the larger point-spread function [PSF] in F444W as the photometric apertures were defined on F356W).  We apply these simulation-based corrections (comparable to similar corrections applied in {\it HST} studies; \citealt[e.g.][]{finkelstein22}) to all fluxes and uncertainties to complete our total flux measurements.  All fluxes and uncertainties were corrected for Galactic attenuation assuming a field-averaged E(B-V)$=$0.006 and a \citet{cardelli89} Milky Way attenuation curve.  We also measure fluxes in a range of circular apertures; as these are used for detection significance tests, we do not correct them to total fluxes (though they are still corrected for Galactic attenuation).

We derive flux uncertainties directly from the data, following \citet{finkelstein22}, based on previous methodology outlined in \citet{papovich16}.  We fit for the noise as a function of aperture size by measuring the fluxes at $\sim$5$\times$10$^{3}$ randomly-placed positions in 15 circular apertures with diameters ranging from 1 -- 100 pixels, fitting a polynomial function to the standard deviation in aperture fluxes as a function of the number of pixels in each aperture.  We then use this function to calculate the photometric uncertainties for each object for a given aperture area.  These values were scaled by the ratio of the error image value at the central position of a given source to the median error value of the whole map.  All aperture and Galactic attenuation corrections were applied to these uncertainties.  
Finally, around each source in our catalog, we calculate a ``local"
noise estimate, as the standard deviation in flux values from the 200
closest of these previously placed random apertures.

\subsection{Photometric Redshifts}

We use the \textsc{EAZY} \citep{brammer08} software package to
estimate photometric redshifts for all sources in our photometric
catalog.  \textsc{eazy} fits non-negative linear combinations of
user-supplied templates to derive probability distribution functions
(PDFs) for the redshift, based on the quality of fit of the various
template combinations to the observed photometry for a given source.
The template set we use includes the ``tweak\_fsps\_QSF\_12\_v3" set
of 12 FSPS \citep{conroy10} templates recommended by the \textsc{eazy} documentation.  To
this we add a set of six additional templates spanning bluer colors
than the FSPS models, as Larson et al.\ (2022, in prep) found that
these improve the accuracy of photometric redshift fits for the
expected blue colors of $z >$ 9 galaxies. 

The new templates were created using stellar population models created
with BPASS \citep{eldridge09}.  To generate bluer rest-UV colors than
the initial set of FSPS templates, we
selected BPASS templates with low metallicities (5\%
solar), young stellar populations (log stellar ages of 6, 6.5, and 7
Myr), and were inclusive of binary stars.
We added an additional set of these models inclusive of nebular
emission lines derived with CLOUDY \citep{cloudy} using a high
ionization parameter (log $U = -$2), a similarly low metallicity, and with
nebular continuum emission. Larson et al.\ (in prep) validated the efficacy of including these
models by testing the recovery of photometric redshifts from a
mock catalog derived by a semi-analytic model \citep{yung22}, finding
that the inclusion of these six additional templates significantly
improved the photometric redshift estimates for blue high-redshift galaxies.

We do not use a luminosity prior (e.g., a flat prior is assumed) as
the epoch in question is completely unexplored, and we include a
systematic error of 5\% of the observed flux values.  Our fiducial
\textsc{eazy} run uses our total fluxes derived from our
Kron-aperture measured colors, and we use the measured fluxes
  and errors even in the case of non-detections (as opposed to using
  upper limits).  We also perform two ancillary runs which we use for
later vetting.  One uses fluxes measured in 0.3\arcs\ circular
apertures (to cover the possibility that a Kron ellipse was drawn
inaccurately, which happens in the presence of bright neighbors).  A
second run had a maximum redshift of $z =$ 7 to allow the exploration
of secondary redshift solutions.

\subsection{Sample Selection}

To select our sample of candidate very high redshift galaxies, we
follow previous work done by our team
\citep{finkelstein15,rojasruiz20,finkelstein22,bagley22}.  We make use
of photometric signal-to-noise criteria, to ensure robust photometric
detections (to minimize the chance of a spurious signal), and ensure
robust non-detections below the Lyman-$\alpha$ break.  We add to these
several criteria based off of the full \textsc{eazy} redshift PDF
(denoted $\mathcal{P}(z)$).  We note that the criteria imposed here
are fairly conservative - we wish to identify the most robust
highest-redshift candidates.  Future work will explore how to relax
some of these criteria to improve sample completeness and push to $z
<$ 12, without introducing unacceptable levels of contamination.

To derive an initial sample of $z \geq$ 12 galaxies, we first impose all following requirements:
\begin{itemize}
    \item Signal-to-noise (SNR) in both F200W and F277W $>$ 6 in
      conservatively small 0.2\arcs\ (6.7-pixel) diameter apertures
      for these measurements, using both the fiducial (global) and
      local noise estimates. We note that enforcing a SNR cut
        in F200W effectively limits this sample to $z \lesssim$ 15,
        though by enforcing detections in both a short and
        long-wavelength detector, we limit inclusion of
        detector-specific spurious sources.
    
    \item Error map values $<$ 1000 (indicating coverage by the majority of exposures) in F115W, F150W, F200W, F277W and the detection image.
    
    \item Initial more inclusive photometric redshift cuts of $\int \mathcal{P}(z>8)$ $\geq$ 0.9, $z_{best} >$ 8.5, $\chi^2_{EAZY} <$ 20 (to reject poor \textsc{eazy} fits), and that the $\Delta z =$ 1 integer redshift bin ($z_{sample}$) with the largest integrated $\mathcal{P}(z)$ to be at $z_{sample} \geq$ 9.
    
    \item Objects at $z_{sample} >$ 10 must have SNR $\leq$ 2.0 in F115W, while objects at $z_{sample} >$ 13 must have a SNR $\leq$ 2.0 in both F115W and F150W (in both the global and local noise in 0.2\arcs-diameter apertures).  These redshifts correspond to the wavelength of the Lyman-$\alpha$ break leaving a given dropout filter.  
    
    \item F200W magnitude $<$ 29, to focus on well-detected objects regardless of formal SNR.
\end{itemize}

After this initial set of selection criteria, we examined the resulting objects.  We inspected their spectral-energy distributions (SEDs), image stamps, and $\mathcal{P}(z)$ plots.  We noticed several low-confidence sources which could be identified with further automated cuts.  We thus implemented this additional set of selection criteria:
\begin{itemize}
\item We additionally implement all of the above detection significance criteria, both in the detection and dropout bands, in a 0.3\arcs-diameter aperture to account for situations where faint flux was visible slightly off-center of the source barycenter.

\item We require the $\chi^2$ from an additional \textsc{eazy} run with a maximum redshift of seven to have a significantly worse fit than our fiducial run via $\chi^2_{Low-z} - \chi^2_{fiducial} > 4$.

\item We impose a single color cut of F200W - F444W $<$ 1 for objects
  with $z_{sample} \leq$ 13 to reduce the incidence of red
  low-redshift interlopers.  This is similar to the color cuts
  simulated by \citet{hainline20} and implemented by
  \citet{castellano22}.  For objects with $z_{sample} =$ 13--18 the
  Ly$\alpha$ break falls in the F200W filter, thus we
  require F277W - F444W $<$ 1.

\item To account for situations where the Kron aperture could be affected by nearby bright sources, we also require $\int \mathcal{P}(z>8)$ $\geq$ 0.5 from an independent \textsc{eazy} run performed with colors measured in 0.3\arcs\ circular apertures.

\end{itemize}

As our focus here is on the highest-redshift sources, we limit our
analysis to objects with $z_{sample} \geq$ 12.  Running the above
selection process on all four fields, we find a single galaxy
candidate which satisfies all of the above criteria.  We perform an initial visual inspection of this candidate,
inspecting 1.5\arcs\ image stamps in all filters, and 5\arcs\ image
cutouts in F200W and the detection image, and find that this object
appears astrophysical in origin, and is not an artifact.

\begin{figure*}[!t]
\includegraphics[width=\textwidth]{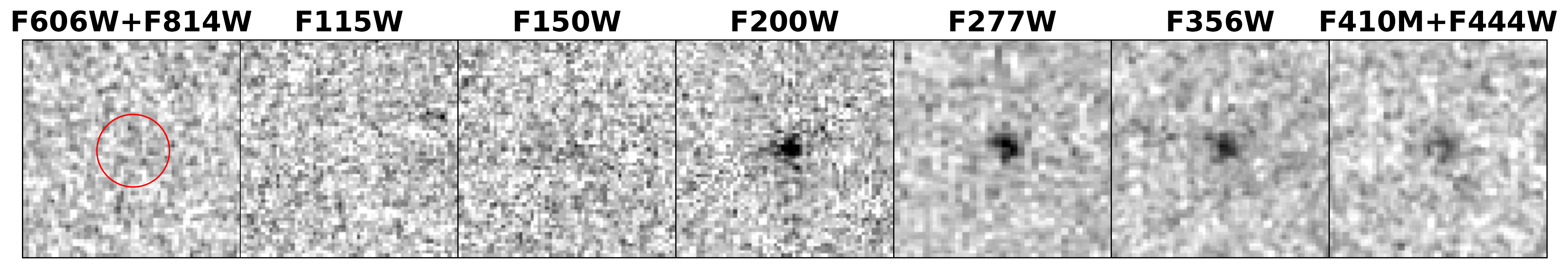}
\includegraphics[width=\textwidth]{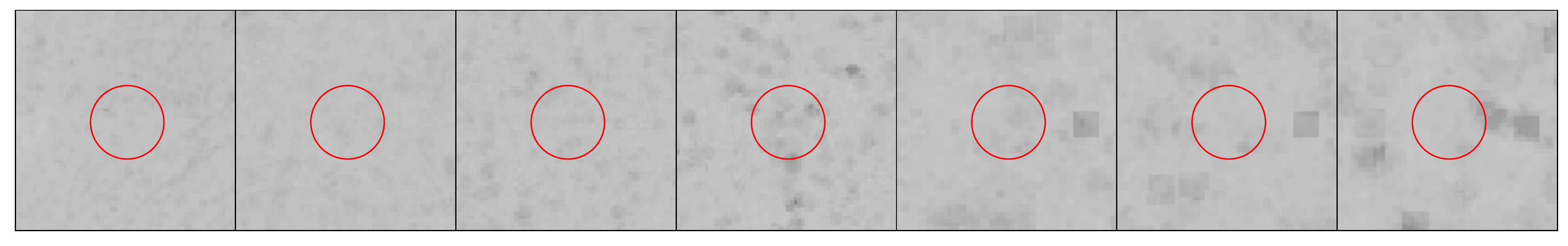}
\vspace{-2mm}
\caption{Top) 1.8\arcs $\times$ 1.8\arcs\ cutout images centered on
  the position of Maisie's Galaxy in the non-PSF-matched images.  This
  source exhibits the hallmark colors of a distant galaxy -- no
  discernible flux in a dropout band (we show stacked F606W$+$F814W as
  well as F115W images; the circle has a radius of 0.3\arcs) and a
  significant detection in the bluest detection band (F200W in this
  case). Very faint flux is visible (at $\sim$2.8$\sigma$ significance
  when measured in a 0.2\arcs\ diameter aperture) in F150W, which
  drives the redshift to $z \sim$ 12. The wide wavelength range of NIRCam allows this source to be well-detected in multiple filters, and in the imaging alone it is clear this source exhibits a blue spectral shape.  Bottom) Same ordering as the top, for sky-uncertainty maps constructed from the variance of the readout noise, all using a linear scale from 0.33 to 3$\times$ the robustly-measured sky standard deviation in each band. The patchiness of the uncertainties is due to loss of exposure time when cosmic-rays are detected and rejected in the multiple readouts or in outlier rejection when combining the dithered exposures (the 2$\times$ larger original pixel scale of the long-wavelength channels results in larger patches than the short-wavelength channels).  The uncertainty arrays show no excess in rejected pixels near the candidate galaxy.}
\label{fig:stamps}
\end{figure*}

\begin{deluxetable}{lc}
\vspace{2mm}
\tabletypesize{\small}
\tablecaption{Properties of Maisie's Galaxy}
\tablewidth{\textwidth}
\tablehead{\multicolumn{1}{c}{Property} & \multicolumn{1}{c}{Value}}
\startdata
Source ID&CEERSJ141946.36$+$525632.8\\
RA (J2000 [deg])&214.943153\\
Dec (J2000 [deg])&52.942442\\
$z_{EAZY}$ & 11.8$^{+0.2}_{-0.3}$\\
$\mathcal{T}_{Big Bang}$ & 373$^{+16}_{-8}$ Myr \\
\hline
$M_{UV}$ (mag)& $-$20.32$^{+0.08}_{-0.06}$ \\
$\beta$ & $-$2.47$^{+0.09}_{-0.09}$\\
log (M$^{\ast}$/M$_{\odot}$) & 8.50$^{+0.29}_{-0.44}$ \\
A$_{v}$ (mag) & 0.07$^{+0.23}_{-0.06}$ \\
SFR$_{10 Myr}$ (M$_{\odot}$ yr$^{-1}$) & 2.1$^{+4.8}_{-2.0}$ \\
log sSFR$_{10 Myr}$ (yr$^{-1}$) & $-$8.2$^{+1.0}_{-1.6}$ \\
Mass-weighted Age (Myr) & 18$^{+18}_{-9}$
\enddata
\tablecomments{$\mathcal{T}_{Big Bang}$ is the time elapsed from the
  Big Bang to the photometric redshift for our assumed cosmology.
  M$_{UV}$ and $\beta$ were computed from the Prospector models, using the same techniques as in
\citet{tacchella22}.  The physical properties listed below the horizontal line were derived with \textsc{Prospector}.}
\label{tab:properties}
\vspace{-8mm}
\end{deluxetable}

%Mstar: 8.521 8.26 8.772
%Av=0.079 0.019 0.247
%log SFR_10= 0.254 -1.192 0.759
%log sSFR 10 = -8.311 -9.937 -7.574
%age: 0.019 0.015 0.037
%met: -1.722 -1.919 -1.412

\section{Results}

\subsection{A Robust Galaxy Candidate at $z \approx$ 12}
This source, CEERSJ141946.36+525632.8, hereafter known as ``Maisie's
Galaxy"\footnote{This exceptional source survived all detailed
  analysis steps, firmly becoming a plausible candidate on the ninth
  birthday of the lead author's daughter. We adopt this short name for
  convenience in this and future papers.} was detected in the CEERS2
field.  Promisingly, it was first identified in the earliest (v0.02)
internal CEERS reduction in this field, being the first $z \gtrsim$ 12
candidate viewed on 18 July, 2022.  In each subsequent reduction, this
source continued to satisfy all selection criteria, becoming
progressively more robust as the data became cleaner.  Several CEERS
team members viewed this potential $z \gtrsim$ 12 candidate
on 22 July 2022, and agreed on the robustness of this source.  The
photometric redshift of this source with NIRCam photometry alone is $z =$ 12.0$^{+0.2}_{-0.4}$.

Given the much greater sensitivity of NIRCam, we do not expect
significant changes to our interpretation based on the inclusion of
{\it HST} images. These images were not included in our fiducial
\textsc{SE} analysis as they had not yet been pixel aligned given the
short time since NIRCam data acquisition.  However, upon inspection we
find a hint of a positive signal at the position of the source in the
F160W image. Indeed, while the source is not in in the published
\citet{stefanon17} and \citet{skelton14} catalogs, there is a $3.5
\sigma$ detection at a separation of 0.15\arcs\ in the
\citet{finkelstein22} catalog (this object has SNR $<$ 2 in all other
{\it HST} filters in this catalog).

Using \textsc{SE} we perform forced photometry at this position on the
CANDELS \citep[e.g.,][]{grogin11,koekemoer11} 30 mas images (using the
updated images provided by the CEERS team which have registered the
astrometry to {\it Gaia}) in the ACS F606W and F814W, and WFC3 F125W,
F140W and F160W bands.
Following \citet{finkelstein22}, we apply an
additional flux correction of 1.2 to all {it HST} bands to account for
missing wings of the PSF in the larger Kron aperture.  We find SNR
$<$2 in all bands except WFC3 F160W, which has a $\sim$4$\sigma$
detection, with a flux consistent with the weak F150W detection for
this source.  Including this photometry (listed in Table
3) in our photometric redshift fit slightly changes our photometric
redshift estimate to $z =$ 11.8$^{+0.2}_{-0.3}$,
which (for our assumed cosmology) corresponds to an age of the
Universe of 373$^{+16}_{-8}$ Myr.  We use these results inclusive of
{\it HST} photometry as our fiducial values.  The properties of this galaxy are
summarized in Table~\ref{tab:properties}, and we list its photometry
in Tables~\ref{tab:tab2} and ~\ref{tab:tab3}.

\begin{figure}[!t]
\epsscale{0.55}
\plotone{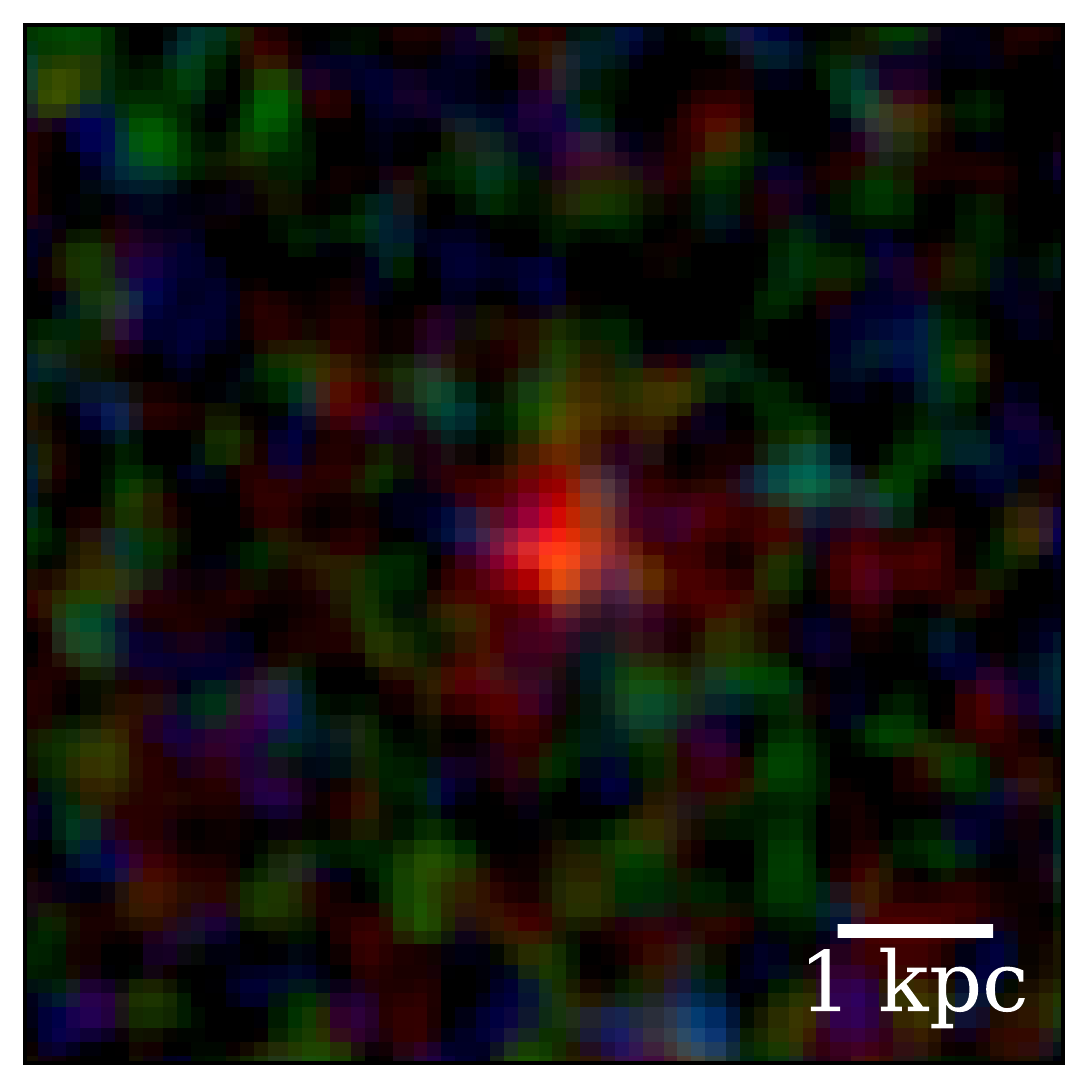}
\plotone{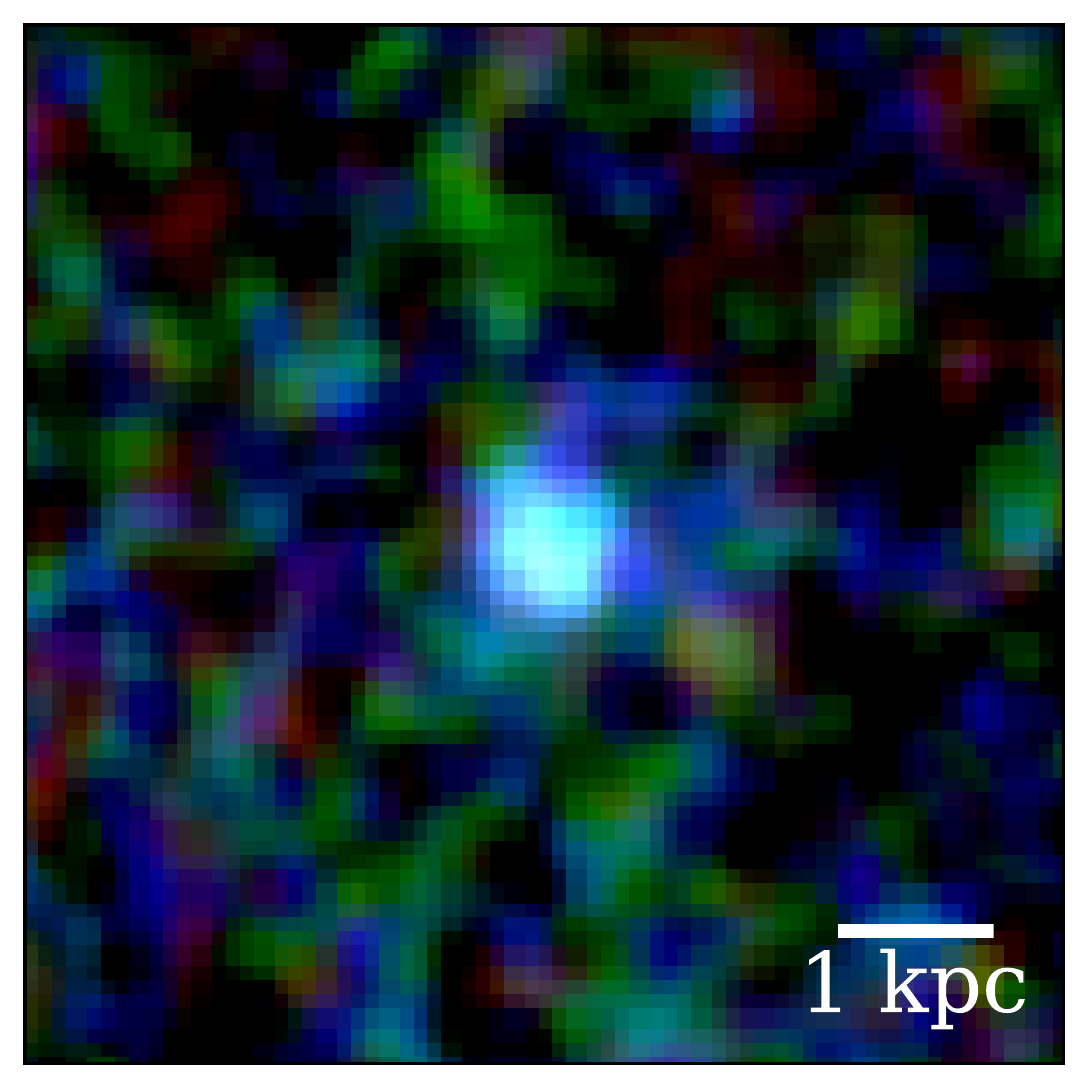}
\caption{Three-color images of Maisie's Galaxy.  The left image is a
  composite of {\it HST}/ACS F606W and F814W in blue, F115W and F150W
  in green, and F200W in red.  This shows the galaxy candidate as red
  due to the very high redshift resulting in no detected flux in the
  filters assigned to the blue and green colors.  The right image
  shows an approximated ``true" rest-UV color image, composed of
  F200W$+$F277W in blue, F356W in green, and F410M+F444W in red.  As we discuss further in \S 5, intrinsically this galaxy is quite blue.  The scale bar corresponds to 1~(physical) kpc assuming $z = 12$ at a scale of 0.37\arcs\ per kpc. }
\label{fig:color}
\end{figure}

\begin{figure*}[!t]
\epsscale{1.2}
\plotone{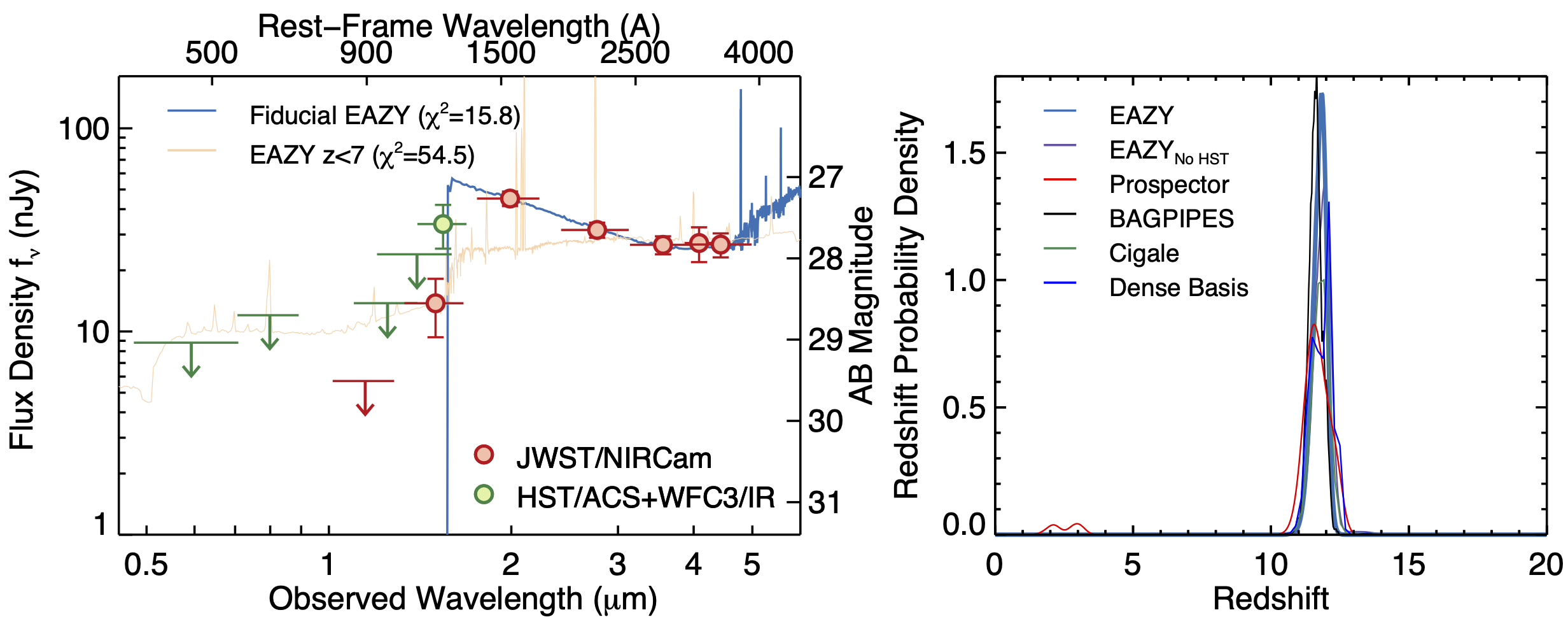}
\caption{Left) The circles denote our fiducial photometry, with green
  and red denoting {\it HST}/ACS$+$WFC3 and NIRCam instruments,
  respectively.  This SED exhibits the hallmark shape of a
  high-redshift galaxy, with several non-detections in blue filters,
  followed by significant detections with a blue spectral slope.  The
  arrows denote 1$\sigma$ upper limits.  The F115W-F200W break color
  is $>$1.9 mag (2$\sigma)$, which is sufficient to rule out all
  low-redshift solutions. The blue curve shows the best-fitting
  \textsc{eazy} model at $z =$ 11.8.  The orange curve shows the
  result if we force \textsc{eazy} to find a solution at $z <$ 7.
  This model is unable to match the amplitude of the break as well as
  the slope redward of the break, and is correspondingly ruled out at
  high confidence ($\chi^2_{low-z} =$ 54.5, compared to 15.8 for the $z
  =$ 11.8 solution).  Right) Photometric redshift probability
  distribution functions for Maisie's Galaxy.  The thick blue curve
  shows the fiducial PDF from \textsc{eazy}, which exhibits no
  low-redshift solution and a peak at $z =$ 11.8$^{+0.2}_{-0.3}$ (the
  purple curve shows the \textsc{eazy} result without {\it HST}
  photometry, which prefers $z =$ 12.0 and has a tail to $z =$ 14).  The
  remaining curves show the results from independent runs with
  \textsc{Prospector}, \textsc{Bagpipes}, \textsc{Cigale} and
  \textsc{Dense Basis} (see \S 5).  All results significantly prefer a
  $z >$ 12 solution, with all four codes finding best-fit redshifts
  nearly identical to \textsc{eazy} ($z =$ 11.8$^{+0.2}_{-0.3}$, $z =$ 11.6$^{+0.2}_{-0.2}$, $z =$
11.8$^{+0.4}_{-0.4}$, and 11.9$^{+0.4}_{-0.4}$, respectively).}
\label{fig:sedpz}
\end{figure*}

Figure~\ref{fig:stamps} shows cutouts of this candidate galaxy in the
NIRCam bands, while Figure~\ref{fig:color} shows two color composites.
Figure~\ref{fig:sedpz} shows the observed spectral energy distribution
of our candidate with photometric redshift fits.  The confidence of
this source as a robust very high-redshift galaxy is easy to see from
all three of these figures.  The Lyman-$\alpha$ break color, here
F115W-F200W, is $>$1.9 mag (2$\sigma$ lower limit), completely
eliminating any known low-redshift interloper (the F150W-F200W color
is 1.3 mag; still strong, though smaller in amplitude due to the
Ly$\alpha$ break being present at the very red edge of F150W).  Such
a model would need to have an extremely red color to match our
F115W-F200W $>$1.9 mag break, but then have a very blue color.  While
lower-redshift passive or dusty galaxies can mimic high-redshift
Lyman-$\alpha$ breaks, the observed $>$1.9 mag break is much larger
than known populations of low-redshift galaxies.  Such galaxies would
also be fairly red redward of the break.  

Though differential geometry could accommodate UV spectral slopes as
blue as $\beta \sim -$ 1, this object has $\beta \sim -$2.5 \citep[see
\S 5.1;][]{casey14b}.  The significant detection in four broadband
filters also rules out low-redshift extreme emission line galaxies. We
show as the orange curve in Figure~\ref{fig:sedpz} \textsc{eazy}'s
best-fitting low-redshift model, which is ruled out at high
confidence.  Based on the non-detection in F115W and strong detection
in F200W, the implied redshift is $z >$ 11.  This is confirmed by the
\textsc{eazy} fit, shown as the blue line, which prefers $z \sim$ 11.8.

\subsection{Fidelity of Candidate}\label{sec:fidelity}

Figure~\ref{fig:stamps} shows 1.8\arcs\ cutout images of this source
at multiple wavelengths.  This source shows the expected pattern for a
high-redshift galaxy, with no significant flux in multiple dropout
bands, with robust flux in redder bands.  The very sharp break between
F115W and F200W is consistent with a redshift of $z >$ 11 (the faint
detection in F150W pushes the redshift solution to $z \sim$ 12 as
opposed to higher redshift).  The advantage of {\it JWST} is clear here, as this source is well-detected in all five NIRCam filters redward of the break.  This multi-band detection essentially eliminates the possibility of a spurious nature.  Of note is that while persistence from previous observations affected several {\it HST} programs (see discussion in \citealt{finkelstein22} and \citealt{bagley22}), CEERS observed with the bluest filters first, thus any flux from persistence would be most apparent in F115W.

To further rule out a spurious nature, the science, error, and data-quality images were visually inspected at the position(s) of the best candidate(s). This is to ensure that the detected sources in the co-added images are not just chance super-positions of regions that were affected by cosmic rays or other artifacts. In the case of the $z \sim 12$ candidate, the source is visible in all of the individual F200W, F277W, and F356W exposures, and overlaps with a cosmic ray in only a few images. Even in those cases, the cosmic rays that are masked in the data-quality array are of the typical size that is cleanly rejected in the jump-detection step of the pipeline. There were no overlaps with the larger ``snowball" charged-particle events. 

As an additional check, we measured photometry at the position of this source on our images without our post-processing residual background subtraction step, to ensure any systematic effects at this source position did not affect our results.  The images already have a pedestal background subtracted in the pipeline, so the relative colors should be secure when measured in this way.  We found that this set of photometry was consistent with our fiducial photometry, and \textsc{eazy} returns $\int \mathcal{P}(z>11) =$ 0.995.  Therefore it is unlikely that our sky subtraction routine negatively affected our result.

Dust-reddened foreground galaxies are another potential source of contamination.  However, Maisie's Galaxy is not significantly detected in the deepest mid- and far-infrared, sub-millimeter and radio data available for this sky region, including  Spitzer MIPS 24$\,\mu$m \citep{magnelli09}, Herschel PACS 100$\,\mu$m and 160$\,\mu$m \citep{lutz11},
Herschel SPIRE 250$\mu$m, 350$\mu$m, and 500$\mu$m \citep{oliver12}, JCMT SCUBA2 850$\mu$m \citep{geach17}, 
and VLA 10~cm (Dickinson, priv.\ comm.).  The SCUBA2 non-detection is
explored in more detail in \citet{zavala22}.

\begin{deluxetable*}{ccccccc}
\vspace{2mm}
\tabletypesize{\small}
\tablecaption{Measured Photometry of Maisie's Galaxy with {\it JWST}/NIRCam}
\tablewidth{\textwidth}
\tablehead{
\colhead{F115W} & \colhead{F150W} & \colhead{F200W} & \colhead{F277W} & \colhead{F356W} & \colhead{F410M} & \colhead{F444W}
}
\startdata
$-$7.52 $\pm$ 3.80&13.77 $\pm$ 4.40&45.11 $\pm$ 3.69&31.62 $\pm$ 2.68&26.29 $\pm$ 2.70&27.26 $\pm$ 5.30&26.78 $\pm$ 3.61\\
\enddata
\tablecomments{Fluxes are in nJy, and \edit1{correspond to total fluxes}.  AB magnitudes can be derived via: $-$2.5 log$_{10}$ (f$_{\nu}$[nJy]) $+$ 31.4.}
\label{tab:tab2}
\end{deluxetable*}

\begin{deluxetable}{ccccc}
\vspace{2mm}
\tabletypesize{\small}
\tablecaption{Measured Photometry of Maisie's Galaxy with {\it HST}}
\tablewidth{\textwidth}
\tablehead{
\colhead{F606W} & \colhead{F814W} & \colhead{F125W} & \colhead{F140W} & \colhead{F160W}
}
\startdata
7.3 $\pm$ 5.9&4.8 $\pm$ 8.0&-7.8 $\pm$ 9.2&-37.2 $\pm$ 16.0&33.8 $\pm$ 8.2\\
\enddata
\tablecomments{Fluxes are in nJy, and \edit1{correspond to total fluxes}.  AB magnitudes can be derived via: $-$2.5 log$_{10}$ (f$_{\nu}$[nJy]) $+$ 31.4.}
\label{tab:tab3}
\end{deluxetable}

\subsubsection{Stellar Screening}
Low-mass stars and brown-dwarfs can have colors that mimic
high-redshift galaxies in broadband filters
\citep[e.g][]{Yan03,Ryan05,Caballero08,Wilkins2014} in the absence of
longer wavelength observations ($\lambda_{\rm obs}\!\gtrsim\!2~\mu$m).
We explore this possibility following the methodology in
\citet{finkelstein22}.  In brief, we derive a grid of models for the
colors of low-mass stars and brown dwarfs (spectral types of M4--T8)
in the NIRCam filters, by integrating the IRTF SpEX brown dwarf
templates \citep{burgasser14}.  As these spectra end at 2.5$\mu$m, we
use the tabulated 2MASS photometry to link each SpeX model with {\it
  Spitzer}/IRAC photometry from \citet{patten06}.  As the differences
in filter transmission are negligible, we assume we can map IRAC
3.6$\mu$m onto F356W and 4.5$\mu$m onto F444W, however this assumption
will need to be revisited with future spectroscopic observations of
brown dwarfs with {\it JWST} at $\lambda\gtrsim2.5~\mu$m. We estimate
the best brown dwarf template would be an L7.5-dwarf, and such a source
would have blue near-infrared color of F115W$-$F200W $=$ 0.9
mag. This is strongly ruled out by our observation of
F115W$-$F200W$>$1.9 mag (2$\sigma$ lower limit).  Additionally, our
size analysis in \S 5.2 shows that this object is inconsistent with a
point source.

\subsubsection{Photometric Accuracy}
While our fiducial photometric measurements were derived in as robust a manner as possible, different software packages require different parameters and assumptions, which could lead to unknown systematic biases.  We thus independently derive NIRCam photometry from our images with two independent software packages.  The first method is \textsc{Photutils} from Python's astropy package \citep{photutils}. Source detection was performed on a combined F277W and F356W image and the resulting segmentation image passed to the \textsc{Photutils} SourceCatalog routine, which carried out aperture-matched photometry on the background-subtracted, PSF-matched images in each filter.

The second method is a custom photometry package, where photometry is measured in circular apertures with radii ranging from 0.10\arcs\ to 0.35\arcs, applying aperture corrections for point-like sources ($<$0.1~mag for r$>$0.25\arcs), and after locally (30\arcs\ box) aligning the images \citep{perezgonzalez08}. Sky noise measurements in a 6\arcs\ $\times$ 6\arcs\ box around the source take into account correlated noise and are used to quote 5$\sigma$ upper limits for non-detections. Photometric differences for each band are smaller than 0.1~mag for apertures between 0.2\arcs\ and 0.35\arcs, 0.3-0.6 mag fainter for smaller radii, indicating that the source is (slightly) resolved.  This method was applied to the non-PSF-matched imaging.

Comparing results between our fiducial \textsc{SE} photometry and
these independent methods, we find that both the Lyman break and
rest-UV colors show extremely high consistency.  
The upper limits in F115W are similar to our fiducial values.  The
F115W$-$F200W Lyman-$\alpha$ break color is $>$2.3 mag (2$\sigma$)
from the custom method, and $>$1.7 mag (2$\sigma$) for
\textsc{Photutils} (compared to $>$1.9 mag [2$\sigma$] for our
fiducial \textsc{SE} photometry).  Comparing colors, our measured F200W$-$F444W color of
$-$0.6 $\pm$0.4 is highly consistent with the measurement from the
custom method of  $-$0.7$\pm$0.3.  The \textsc{Photutils} measurement
is even bluer ($-$1.0 $\pm$ 0.4) due to a 0.4 mag fainter F444W
measurement.  
 We conclude that while differences in
photometric packages and associated assumptions (in particular
aperture corrections) can affect the
photometry at the $\sim$10-30\% level in most cases, this does not
affect the validity of our candidate as these independent methods find
a consistently strong Lyman-$\alpha$ break followed by a blue spectral
slope, fully consistent with our interpretation of a $z \sim$ 12 galaxy.

\subsubsection{Photometric Redshift Accuracy}
Similar to photometry, different photometric redshift packages can also impart biases on results.  While we have used a well-tested fiducial package in \textsc{eazy}, and implemented a new set of templates customized for very high-redshift galaxies, it is prudent to explore whether other packages would find different photometric redshift results.  As we discuss below, we have run the \textsc{Prospector} \citep{prospector}, \textsc{Bagpipes} \citep{carnall18}, \textsc{Cigale} \citep{burgarella05,noll09,boquien19} and \textsc{Dense Basis} \citep{iyer17,iyer19} SED-fitting codes on our fiducial photometry.  While for the stellar population properties discussed below, we use our \textsc{eazy}-derived $\mathcal{P}(z)$ as a redshift prior, we also performed an independent run with the redshift as a free parameter.  Figure~\ref{fig:sedpz} shows our fiducial \textsc{eazy} $\mathcal{P}(z)$ along with the redshift PDFs from these independent runs.  

These five results show remarkable consistency, all preferring $z >$
12 with no significant low-redshift solutions. All four codes find
results simular to our fiducial \textsc{eazy} run.
\textsc{Prospector} finds $z =$ 11.8$^{+0.2}_{-0.3}$,
\textsc{Bagpipes} finds $z =$ 11.6$^{+0.2}_{-0.2}$, \textsc{Cigale}
finds $z =$ 11.8$^{+0.4}_{-0.4}$, and \textsc{Dense Basis} finds
$11.9^{+0.4}_{-0.4}$.  Combining the posteriors of all four
photometric redshift estimates provides a redshift PDF in agreement
with our fiducial \textsc{eazy} results with a median redshift of
11.74, and a 97.5\% confidence that $z >$ 11.0.  We conclude that
systematic biases due to choices in photometric redshift analyses are
not affecting our results.  Our fiducial result uses that from
\textsc{eazy} as it used templates trained on observations, while the
full grids spanned by the other four codes may include unphysical
parameter combinations.

As one final test, we explore the impact of our inclusion of the six
additional blue templates in our photometric redshift analysis.
Refitting our photometry with the standard templates we find $z
=11.50_{-0.33}^{+0.33}$, consistent within $1\sigma$ of our fiducial
result.  However, standard templates produce a significantly worse fit
($\chi^2 \sim$ 34 versus $\sim$15 for the fiducial fit).  The standard
best-fitting template has a F200W-F277W color of $-$0.04 mag, while
our fiducial template has F200W-F277W=$-$0.35; both can be compared to
the observed color of F200W-F277W=$-$0.39.  It is clear that our
inclusion of bluer templates are better able to match the colors of ultra
high-redshift galaxies such as the one we present here.

\begin{figure*}[!t]
\epsscale{1.2}
\plotone{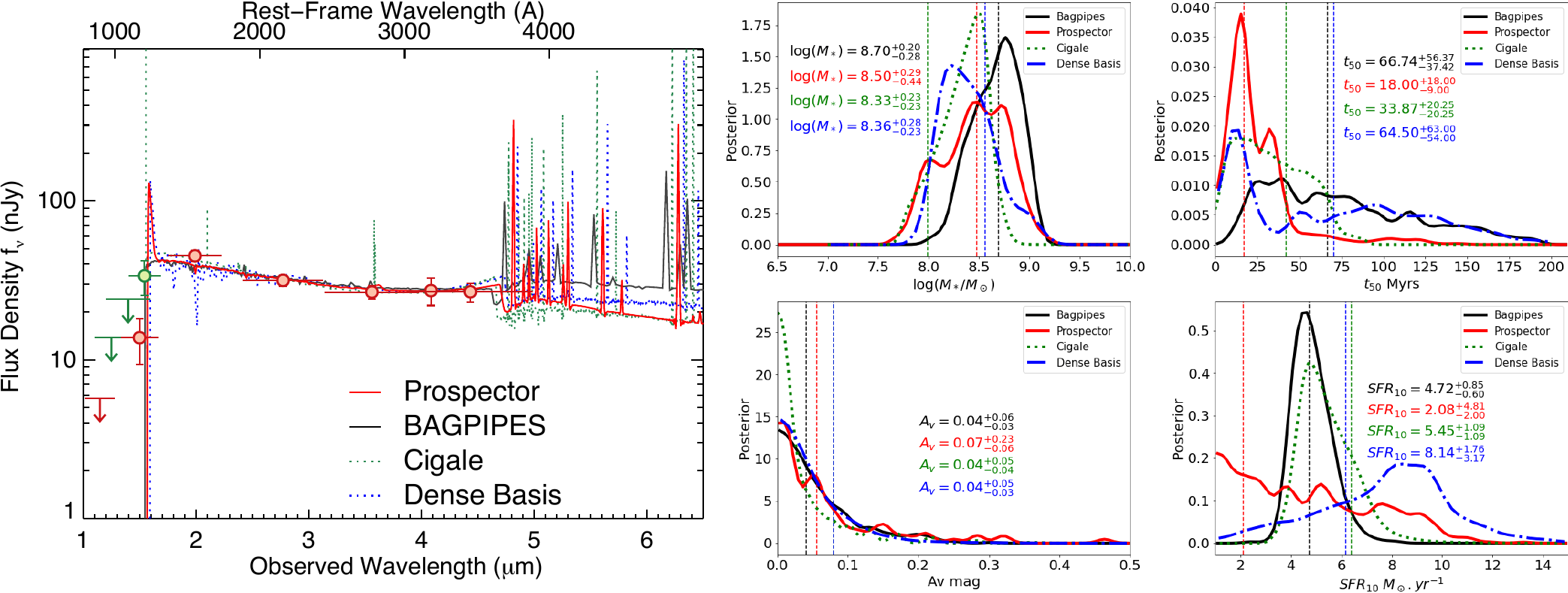}
\caption{Left) Plot showing our fiducial photometry of Maisie's Galaxy alongside best-fit SED models from the SED fitting code \textsc{Prospector} (red line; fiducial, see Table~\ref{tab:properties}), \textsc{Bagpipes} (black), \textsc{Cigale} (green dotted) and \textsc{Dense Basis} (blue dotted). Right) Posterior distributions of the key stellar population properties from all four codes. The panels show stellar mass, mass weighted age, dust attenuation and SFR averaged over the last 10 Myrs. The vertical dotted lines indicate the mean of the posteriors. Posteriors of attenuation are consistent between all four codes.  \textsc{Prospector} prefers a younger age than the other three because of a recent burst in the SFH of this object. As a consequence of the burst, \textsc{Prospector} also estimates a lower SFR. The four estimates of the stellar mass posteriors exhibit significant overlap, though the median values differ by $\pm$ 0.2 dex.  Future observations in the rest-optical with MIRI could break these degeneracies.}
\label{fig:stellarpops}
\end{figure*}

\subsubsection{Contamination Estimation}
To determine the likelihood that our selection criteria would produce a low-redshift contaminant we imposed our same selection criteria cuts on the simulated catalogs used for all the mock CEERS observations. We note that there are zero $z>10$ sources in this catalog so recovery of any source using these selection criteria would indicate contamination of our high-redshift sample. More information about the simulation used can be found in \citet{yung22} and \citet{somerville21}. We use the perturbed fluxes as described in Larson et al. (2022, in prep) which use the same method as determined by Bagley et al. (2022, in prep) where they modeled the noise in simulated {\it JWST} images to have a Voigt profile distribution. We used the 1$\sigma$-depth in each filter for our errors and ran the whole catalog through \textsc{eazy}. As our catalog-level fluxes do not have aperture-specific fluxes we cannot impose criteria based on those fluxes.  We apply the following selection criteria to the simulated catalog: SNR in both F200W and F277W $>$ 6, $\int \mathcal{P}(z>8)$ $\geq$ 0.9, $z_{best} >$ 8.5, $\chi^2_{EAZY} <$ 20, SNR $\leq$ 1.5 in F606W \& F814W \& F115W \& F150W, F200W magnitude $<$ 29, F200W-F444W color $<$ 1, and $\chi^2_{Low-z} - \chi^2_{fiducial} > 4$.  Finally, matching the values to those of Maisie's Galaxy, which exhibits SNR $>$ 10 in both F200W and F277W and $\int \mathcal{P}(z>11)$ $\geq$ 0.99, we find zero sources that meet our criteria. This provides further evidence that Maisie's Galaxy has a high-redshift nature. 

\section{Discussion}

\subsection{Physical Properties}

The five photometric detections afforded by NIRCam allow us the
unprecedented opportunity to study the physical properties of a galaxy
potentially only $\sim$400 Myr after the Big Bang. Our fiducial
stellar population modeling is done with the \textsc{Prospector}
Bayesian SED fitting code \citep{prospector}. We follow the same
procedures as in \citet{tacchella22} and we refer the reader there for
more details. Briefly, we model the SED with a 13-parameter model that
includes redshift (where here prior is set to the posterior of
\textsc{Eazy}, unline in \S 4.2.3 where it was free), stellar mass, stellar and gas-phase metallicities, dust attenuation (two-component dust model including birth-cloud dust attenuating young stars ($<10~\mathrm{Myr}$) and nebular emission,  a diffuse component for the whole galaxy with a flexible attenuation law; 3 parameters), and an ionization parameter for the nebular emission. We adopt a flexible SFH prescription with six time bins (the first two lookback time bins are spaced at $0-5~\mathrm{Myr}$ and $5-10~\mathrm{Myr}$, while the other four are log-spaced out to $z=20$; five free parameters) and with the bursty-continuity prior. Furthermore, we assume the MIST stellar models \citep{choi17} and a \citet{chabrier03} IMF. 

To explore how robust these properties are, we perform an independent
fit with the Bayesian \textsc{Bagpipes} \citep{carnall18},
\textsc{Cigale} \citep{burgarella05,noll09,boquien19} and
\textsc{Dense Basis} \citep{iyer17, iyer19} SED-fitting codes. For
\textsc{Bagpipes} we assumed a simple exponential star formation
history with a Chabrier IMF, a Calzetti dust attenuation law and
included nebular emission with an ionization parameter of $10^{-3}$,
with \citet{bruzual03} stellar population models. For \textsc{Cigale},
we assume a delayed star formation history after checking that adding
a burst does not significantly modify the results. \citet{bruzual03}
models with a Chabrier IMF and was used.  \textsc{Dense Basis} was run
using the flexible non-parametric SFH model and priors described in
\citet{iyer19}, assuming a Calzetti dust law and a Chabrier IMF.
Metallicites in all four codes were allowed to vary to allow
  the uncertainty in metallicity to be included in the uncertainties
  on other parameters, though this parameter is not well-constrained
  with photometry alone.  We note that \textsc{Prospector} and
  \textsc{Cigale} assume an error floor (of 5\% and 10\% of the flux,
  respectively), and that \textsc{Prospector}, \textsc{Dense Basis}
  and \textsc{Bagpipes} always fit the measured fluxes, while
  \textsc{Cigale} uses the flux errors as upper limits (by setting the
  flux and uncertainty equal to the measured flux error) when the fluxes are negative.

The marginalized posterior values of the inferred physical properties
from \textsc{Prospector} are summarized in Table~\ref{tab:properties}
and Fig.~\ref{fig:stellarpops}. We infer a stellar mass of
$\log(\mathrm{M}_{\ast}/\mathrm{M}_{\odot})=8.5^{+0.3}_{-0.4}$. The
attenuation in this galaxy is rather low with
$\mathrm{A}_{V}=0.07^{+0.22}_{-0.06}~\mathrm{mag}$, though we stress
that this is not well constrained because we only fit the rest-UV and
it is degenerate with the slope of the attenuation law (which is
variable in this fit). However, the low dust attenuation is in
agreement with the measured UV spectral slope $\beta =
-$2.47$^{+0.09}_{-0.09}$ (measured using the same techniques as in
\citealt{tacchella22}).  This blue color implies little dust, though
does not require extremely low metallicities
\citep[e.g.][]{finkelstein12a,dunlop13,bouwens14}.  Interestingly,
this galaxy is about as blue as $z \sim$ 7 galaxies of similar mass
\citep{finkelstein12a}, implying little evolution in chemical
enrichment between these two epochs for similar-mass galaxies.

We infer a $\mathrm{SFR}_{10}$ (average of the past 10
Myr\footnote{Although the $\mathrm{SFR}_{10}$ would be best estimated
  from nebular emission lines, the (F)UV actually also probes such
  short timescales, in particular for bursty star formation
  \citep[e.g.][]{caplar19, velazquez21} expected at these redshifts.})
of $2.1~\mathrm{M}_{\odot}~\mathrm{yr}^{-1}$ and the corresponding
$\mathrm{sSFR}_{10}$ is $10^{-8.2}~\mathrm{yr}^{-1}$. By looking at
the posterior distribution of the SFH, it becomes apparent that the
model for this galaxy had an episode of elevated star formation
$10-20$ Myr ago with a SFR of
$8_{-7}^{+17}~\mathrm{M}_{\odot}~\mathrm{yr}^{-1}$, i.e. the SFR has
been slightly decreasing in the recent 10 Myr. This explains the
mass-weighted age of $18^{+18}_{-9}$ Myr. This is also consistent with
the half-mass formation time of dark matter halos at $z\sim12$ of a
few tens of Myr \citep{tacchella18}.

These \textsc{Prospector}-based posterior distributions are consistent
with the ones from \textsc{Bagpipes}, \textsc{Cigale} and
\textsc{Dense Basis} (see Fig.~\ref{fig:stellarpops}), though the
difference in age is large (age is defined at half-mass time,
$t_{50}$, which is close to the mass-weighted age). \textsc{Bagpipes},
\textsc{Cigale} and \textsc{Dense Basis} prefer higher age values
(although the posterior distributions are also broader) with
$67_{-37}^{+56}$ Myr, 34 $\pm$ 20 Myr, and $64^{+63}_{-54}$ Myr
respectively.  The SFH inferred from \textsc{Dense Basis} shows a
recent burst of star formation in the last $\sim 30$ Myr. The larger
mass-weighted age comes from the long tail of low-level star formation
in the galaxy leading up to the recent burst.  The spread in these
results could be explained by differences in SFHs (e.g.,
non-parametric versus parametric), and also the lack of observational
constraints in the rest-frame optical.  These differences also lead to
\textsc{Prospector} having the lowest  $\mathrm{SFR}_{10}$ value.

Several pre-{\it JWST} studies have focused on inferring SFHs and stellar ages of $z\approx8-10$ galaxies \citep[e.g.][]{hashimoto18, laporte21, stefanon22}. Specifically, \citet{tacchella22} -- using  \textsc{Prospector} with the same bursty continuity prior -- found a diversity of stellar ages, ranging from 10 Myr to 260 Myr, and stellar masses ($10^9-10^{11}~{\rm M}_{\odot}$), with more massive galaxies being older. In particular the galaxies
at $z\approx9-10$ with stellar masses at the higher end and the older
ages ($t_{\rm 50}\approx100~\mathrm{Myr}$) are consistent with being
the descendants of Maisie's Galaxy. Recently, \citet{naidu22} inferred
the properties of two galaxies at $z\approx10.6$ and $z\approx12.4$
(see also \citealt{castellano22}) with \textsc{Prospector} and a
similar setup, allowing us to do a useful comparison. Their two galaxies
have $\log(\mathrm{M}_{\ast}/\mathrm{M}_{\odot})=9.4_{-0.3}^{+0.3}$
and $9.0_{-0.4}^{+0.3}$,
$\mathrm{SFR}_{50\mathrm{Myr}}=12_{-4}^{+9}~\mathrm{M}_{\odot}~\mathrm{yr}^{-1}$
and $7_{-3}^{+4}~\mathrm{M}_{\odot}~\mathrm{yr}^{-1}$, and
$t_{50}=111_{-54}^{+43}~\mathrm{Myr}$ and
$71_{-32}^{+33}~\mathrm{Myr}$, respectively. This is older than what
we infer for our galaxy, though this age difference could be explained
by the stellar mass difference, along with the higher preferred
redshift for Maisie's Galaxy. Importantly, detailed stellar population
analyses of early galaxies will advance significantly with {\it JWST},
in particular when including spectroscopic information.

\begin{figure*}[!t]
\epsscale{0.55}
\plotone{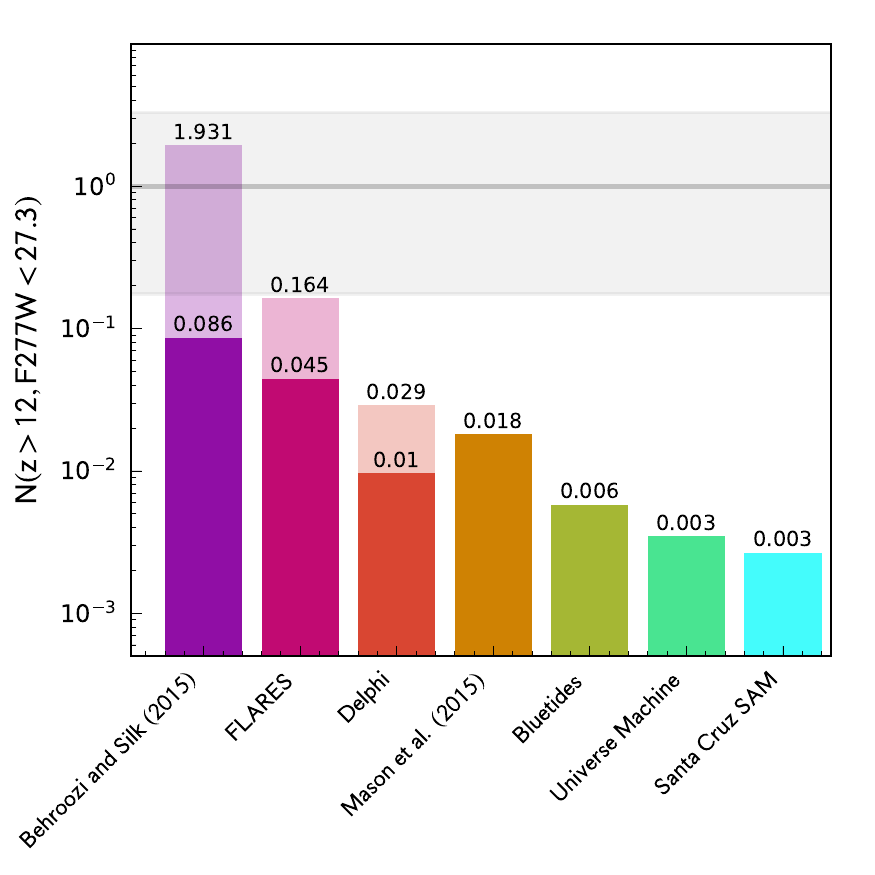}
\plotone{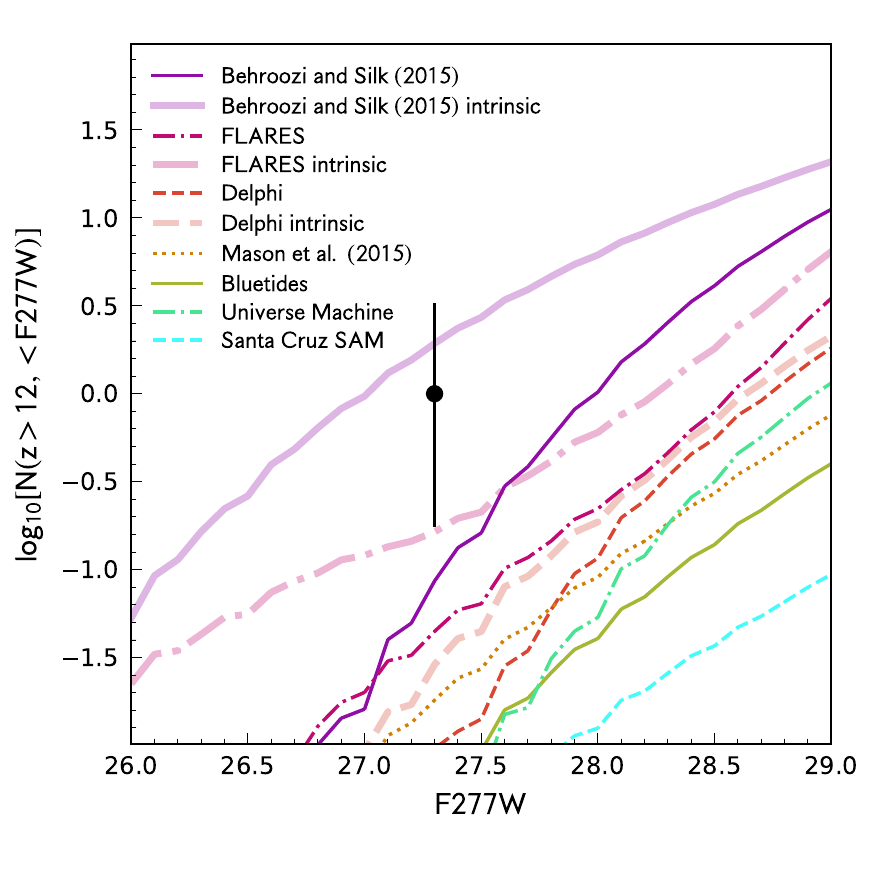}
\vspace{-6mm}
\caption{Theoretical predictions from a range of simulations in the
  recent literature.  The left panel shows the predicted number of
  sources at $m <$ 27.3 (the brightness of our source) and $z >$ 12
  over our survey area of 34.5 sq.\ arcmin.  The vertical axis and the
  values above each bar give the number predicted.  Dark (light)
  shading denotes the value derived from models with (without) dust
  attenuation applied.  The right panel shows these same theoretical
  predictions, now showing the cumulative number as a function of
  apparent magnitude.  The bulk of these models predict that $m \sim$
  27 galaxies at $z \gtrsim$ 12 are not highly likely, though the
  \citet{behroozi15} model, which has no accelerated decline in the
  cosmic SFR density at $z >$ 8, has the least tension.  However, our
  detection of one source has a large Poisson (and cosmic variance)
  uncertainty (gray shading in the left panel; error bar in the right), so strong conclusions cannot yet be made.}
\label{fig:theory}
\end{figure*}

\subsection{Source Morphology}

We derive the sizes of Maisie's Galaxy using two morphological fitting codes, \textsc{GalfitM}\footnote{\href{https://www.nottingham.ac.uk/astronomy/megamorph/}{https://www.nottingham.ac.uk/astronomy/megamorph/}}  \citep{hau2013} and \textsc{statmorph}\footnote{\href{https://statmorph.readthedocs.io/en/latest/}{https://statmorph.readthedocs.io/en/latest/}} \citep{rod2019}. \textsc{GalfitM} is a modified version of \textsc{Galfit}\footnote{\href{https://users.obs.carnegiescience.edu/peng/work/galfit/galfit.html}{https://users.obs.carnegiescience.edu/peng/work/galfit/galfit.html}} \citep{peng2002,peng2010},  a least-squares fitting algorithm that finds the optimum S\'ersic fit to a galaxy's light profile. We perform fits using \textsc{GalfitM} by allowing the S\'ersic index to vary between 0.01 and 8, the magnitude of the galaxy between 0 and 45, and $r_{\textrm{half}}$ between 0.3 and 200 pixels (on our 0.03\arcs\ pixel scale). As input, we use a 100$\times$100 pixel cutout of the F277W science image, the segmentation map created by \textsc{Source Extractor}, and the empirical PSF measured from our CEERS2 pointing, which we allow \textsc{GalfitM} to oversample relative to the data by a factor of nine. We estimate the uncertainty on our fits by conducting a Monte Carlo analysis where we modify the input F277W science image to randomly vary the pixel-to-pixel noise, recompute the parameters, and then repeat this analysis 40 times. 

Following this procedure, we measure a half-light radius of
3.0$\pm$0.12 pixels (0.09 $\pm$ 0.0036\arcs), which corresponds to a
physical size of 340 $\pm$ 14 pc at $z =$ 11.8. We check these results
using the standard configuration of \textsc{Statmorph}, a Python
package developed to calculate the nonparametric morphology of
galaxies as well as compute single S\'ersic fits. Using the same
images as input, we find a half-light radius of 2.9 pixels, in good
agreement with the measurement from \textsc{GalfitM}.  We repeat this
measurement for the F200W filter and a stacked F200W+F277W image and
find consistent results. The measured half-light radius of 3.0$\pm$0.1 pixels is significantly
larger than that expected for a point-source (the median $r_h$ for our
PSF stars is 1.95 $\pm$ 0.22 pixels in F277W), further ruling out a stellar origin for this source.

\subsection{Comparison to Model Predictions}

In Figure \ref{fig:theory} we present predictions from a range of theoretical models, including the First Light And Reionisation Epoch Simulations \citep[FLARES,][]{FLARES-I, FLARES-II, FLARES-V}, a suite of hydrodynamical cosmological zoom simulations; the large periodic volume hydrodynamical simulation Bluetides \citep{Bluetides-I, Bluetides-II}; the Delphi \citep{Dayal2014, Dayal2022} and Santa Cruz SAM \citep{yung19a, yung20b} semi-analytical models, the semi-empirical \textsc{UniverseMachine} \citep{behroozi20}, \citet{Mason2015}, and \citet{behroozi15} models.  For the FLARES, Delphi, and \citet{behroozi15} models, we show both the attenuated and un-attenuated (intrinsic) predictions. These predictions were made by interpolating and integrating either the binned or Schechter luminosity functions across $z=15\to 12$ taking account of the areal size of the CEERS observations. Almost all of these models predict an expected source density much less than one, making the observation of even a single object at this redshift and magnitude surprising and potentially hinting at significant differences between the physical assumptions in these models and the real early universe.  

\begin{figure}[!t]
\epsscale{1.2}
\plotone{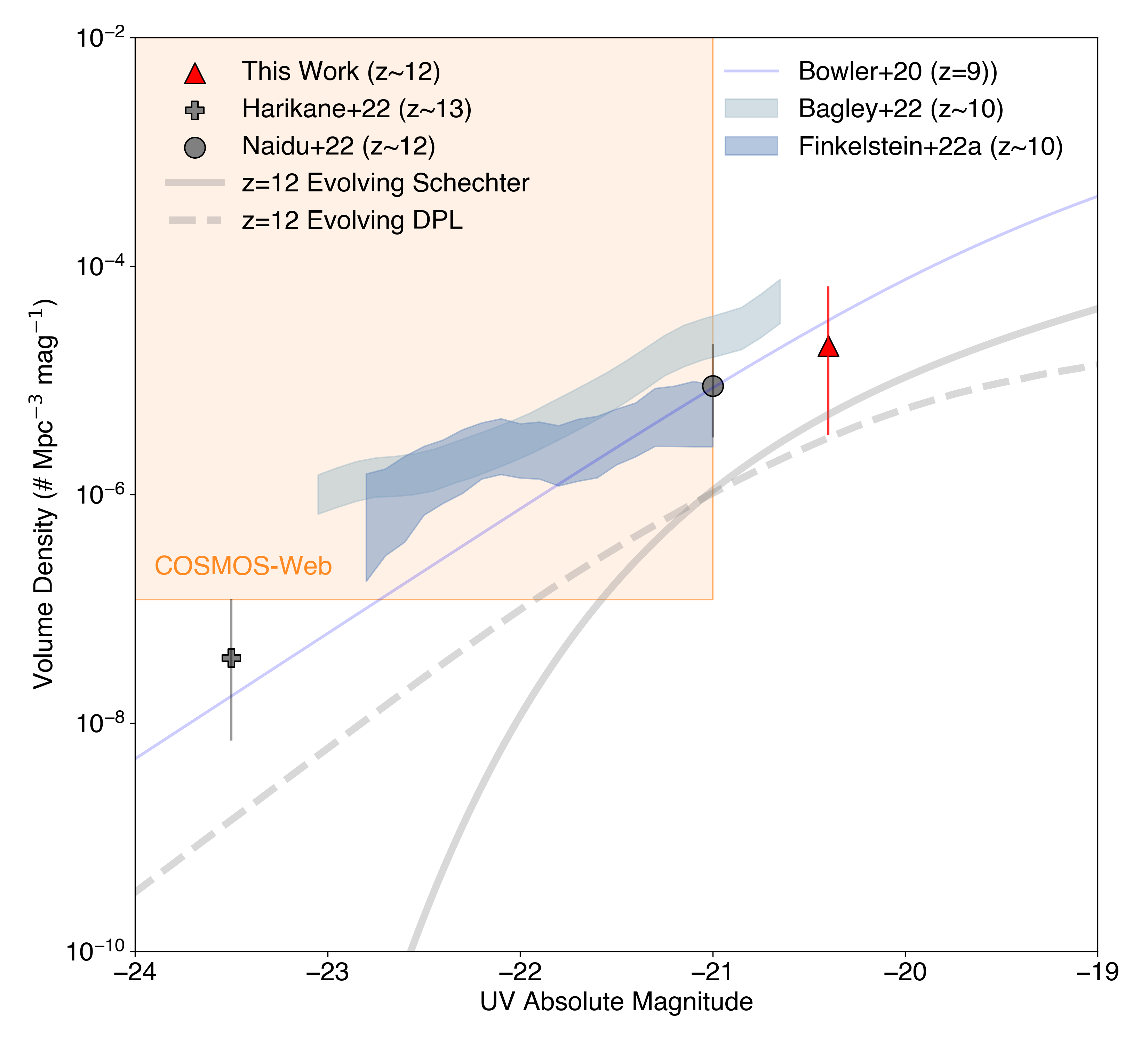}
\caption{A view on the luminosity function at $z \geq$10.  The shaded
  light blue regions show observational constraints at $ z\sim$ 10
  \citep{bagley22,finkelstein22}, while the thin line shows the $z =$
  9 DPL luminosity function from \citet{bowler20}.  The remaining
  points show $z >$ 12 results from this work (triangle), the
  ground-based work of \citet{harikane22}, and the recent {\it JWST}
  work of \citet{naidu22}.  The thick lines show empirical luminosity
  function models which evolve smoothly with redshift, with the solid
  line denoting a Schechter function evolved to $z =$ 12
  \citep{finkelstein16}, and the dashed line a DPL evolved to $z =$ 12
  \citep{finkelstein22b}.  The constraints placed by our observations
  on the faint-end of the luminosity function are consistent with a
  smooth decline out to $z \sim$ 12, though both our observations and
  those of brighter galaxies do lie above the bright-extension of these smoothly-declining functions.  The shaded box shows the parameter-space reached by the upcoming COSMOS-Web survey, which will probe the very bright end at these redshifts.}
\label{fig:z12lf}
\end{figure}

The exception is the \citet{behroozi15} model, which extrapolated galaxy formation to high redshifts by assuming that the ratio between galaxies’ sSFRs and their host halos’ specific accretion rates remained constant, which they showed was equivalent to assuming that galaxies’ stellar masses are proportional to a power of their host halo masses.  This model was constrained only with observational data at $z\le 8$, and predicted no change in the slope of the CSFR relation with redshift at $z>8$.  As a result, it predicted many more high-redshift galaxies than later models that were constrained to match $z\sim 9$ and $\sim 10$ data from \textit{HST} that suggested more rapid declines in the number densities of early galaxies.  We caution against over-interpretation, as the current sample contains only a single object with a consequently large Poisson error in addition to additional uncertainty due to cosmic variance.  Nevertheless, if confirmed, the existence of this object places informative constraints on galaxy formation models in this epoch.

\subsection{Comparisons to Extrapolations from Lower Redshift}
We are now only just getting our first glimpse into this epoch with the first {\it JWST} data. Nonetheless, we can compare our observed number density to a few recent observations.  We calculate a rough number density for $M_{UV}= -$20.3 galaxies assuming a top-hat selection function over 11.5 $< z <$ 12.5.  This is overly simplistic, and does not account for incompleteness (which, although this is a $>$10$\sigma$ detection, certainly is non-unity due to our stringent selection criteria).  Nonetheless it is illustrative of a rough number density.  We find a maximum volume over the CEERS first-epoch area of 5.0 $\times$ 10$^4$ Mpc$^{3}$, for a number density for our singular source of 2.0$^{+4.6}_{-1.7}$ $\times$ 10$^{-5}$ Mpc$^{-3}$ (where the uncertainties are Poisson based on our detection of one object).  

We illustrate this number density in Figure~\ref{fig:z12lf}.  Our
derived number density is not inconsistent with a variety of
observational constraints at $z \sim$ 10, as well as  recent results
at $z \sim$ 12--13.  The solid gray line shows the predicted $z =$ 12
Schechter function from \citet{finkelstein16}, which is extrapolated
from an empirical fit to observations at $z =$ 4--8, assuming smooth
redshift evolution.  Interestingly, our rough number density
measurement is above, though consistent within 1$\sigma$, with this
prediction, which would support its assumption of a smooth decline in
the luminosity function (and in the corresponding star-formation rate
density).  Our results are also consistent with the smoothly-evolving
double-power law (DPL) model at $z
=$ 12 from \citet{finkelstein22b}, again supporting a smooth decline
in the rest-UV luminosity function to $z >$ 10.

As noted by several previous studies the bright-end of the luminosity
function at $z \geq$ 9 exhibits an interesting excess over predicted
levels
\citep[e.g.][]{bowler20,rojasruiz20,morishita18,bagley22,finkelstein22}.
While our survey area does not yet probe the volume densities needed
to reach these brighter potential $z =$ 12 galaxies, if the
high-redshift luminosity function follows a DPL form, the forthcoming
0.6 deg$^2$ COSMOS-Web survey (PIs Kartaltepe \& Casey) should be able
to discover this population.  In combination with the full Cycle 1
slate of surveys, including the completed CEERS imaging, it will
afford a more complete view of the $z =$ 12 universe.

\section{Conclusions}
We present the results from a search for ultra-high-redshift galaxy candidates from the first epoch of NIRCam imaging from the {\it JWST} CEERS survey.  We use imaging from both the short and long-wavelength cameras over four pointings, covering 34.5 sq.\ arcmin in the F115W, F150W, F200W, F277W, F356W, F410W and F444W filters, reaching $m \sim$ 29 (5$\sigma$) in the deepest bands.  We measure photometry using \textsc{Source Extractor}, with an emphasis on robust measurements of colors, total fluxes, and uncertainties.  

We estimate photometric redshifts with the \textsc{EAZY} software package, including new blue templates designed to better recover the colors of very distant galaxies.  We develop iteratively a set of conservative selection criteria to select candidate galaxies at $z \geq$ 12.  We find one candidate galaxy satisfying stringent non-detections (SNR $<$1.5) in all dropout bands, and detected at $>$10$\sigma$ in the detection bands.

This object, dubbed Maisie's Galaxy, has a photometric redshift of 11.8$^{+0.2}_{-0.3}$, and was found in the CEERS2 field.  We explored all known potential sources of contamination, including instrumental effects, systematic biases in the analysis, and contamination by lower-redshift galaxies or Galactic stars.  We find that none of these alternative explanations can account for the observations, leaving us with the conclusion that it is a robust $z \sim$ 12 galaxy candidate.

We explore the physical properties of this unexpected galaxy.  As might be expected for such an early epoch, this galaxy is blue, with a UV spectral slope $\beta = -$2.5, consistent with low levels of dust attenuation.  Stellar population modeling with multiple codes are in agreement that this source has a modest stellar mass of log (M/M\sol) $\sim$8.5, with a high log sSFR of $-$8.2 yr$^{-1}$.  The mass-weighted age of Maisie's Galaxy is young, with a median of $\sim$20 Myr, though stellar populations as old as 100 Myr ($z_{form} \geq$ 14) cannot be ruled out.  The galaxy candidate is significantly resolved, with $r_h =$ 3.0 $\pm$ 0.1 pixels, for a physical size of $\sim$340 pc at $z \approx$ 12.

We compare the abundance of this single galaxy both to model predictions and previous observations.  We find that the presence of this source is unexpected based on most model predictions, though given our sample size the tension is modest at best.  However, both semi-empirical models and empirical extrapolations which assume a smooth decline in the SFR density at $z >$ 8 predict volume densities of such $z \sim$ 12 sources in agreement with our observations.   Should more such sources be found in early {\it JWST} surveys, it would provide further evidence against accelerated decline SFR density scenarios.  

Such a galaxy population would also present challenges for a variety
of dark matter models with suppressed power on small scales, such as
fuzzy dark matter \citep[e.g.][]{sullivan18}, and possibly even for
standard $\Lambda$CDM models. Additionally, the presence of this
galaxy $\sim$370 Myr after the Big Bang may be consistent with
redshifted 21-cm absorption at $z \sim$ 18 reported by the Experiment
to Detect the Global Epoch of Reionization Signature
\citep[EDGES][]{bowman18} presumed to be caused by light from the first stars.

We caution the reader that this galaxy is a candidate.  While we have exhausted multiple avenues to explore whether its presence in our data could be caused by instrumental effects, whether our measurement techniques were biased, or whether its colors could be consistent with lower-redshift sources, the ``gold standard" of distance measurements is spectroscopic confirmation.  Such confirmation should be possible in modest exposure times with the NIRSpec and/or MIRI spectrographs on board {\it JWST}.  The combination of larger samples being compiled by {\it JWST} Cycle 1 programs, including the remainder of CEERS, COSMOS-Web (PIs Kartaltepe \& Casey), JADES (PIs Rieke \& Ferruit), PRIMER (PI Dunlop), PEARLS (Windhorst et al., in prep) and NGDEEP (PIs Finkelstein, Papovich, \& Pirzkal) coupled with subsequent spectroscopic followup will further illuminate the earliest phases of galaxy formation.

\begin{acknowledgements}
We acknowledge that the location where this work took place, the University of Texas at Austin, sits on indigenous land. The Tonkawa lived in central Texas and the Comanche and Apache moved through this area. We pay our respects to all the American Indian and Indigenous Peoples and communities who have been or have become a part of these lands and territories in Texas, on this piece of Turtle Island. 

We thank the entire {\it JWST} team, including the engineers for
making possible this wonderful over-performing telescope, the
commissioning team for obtaining these early data, and the pipeline
teams for their work over the years building and supporting the
pipeline.  The authors acknowledge the Texas Advanced Computing Center
(TACC) at The University of Texas at Austin for providing HPC and
visualization resources that have contributed to the research results
reported within this paper. We thank Brendan Bowler, Caroline Morley,
Jim Dunlop and Mike Boylan-Kolchin for helpful conversations, and
thank the anonymous referee for constructive comments.

We acknowledge support from NASA through STScI ERS award JWST-ERS-1345.  D.\ B.\ and M.\ H.-C.\ thank the Programme National de Cosmologie et Galaxies and CNES for their support.  RA acknowledges support from Fondecyt Regular 1202007. 
\end{acknowledgements}

%% add or uncomment software packages/codes that were used
\software{Astropy \citep{astropy},
          \textsc{Bagpipes} \citep{carnall18},
          \textsc{Cigale} \citep{burgarella05,noll09,boquien19},
          \textsc{Dense Basis} \citep{iyer17,iyer19}, 
          Drizzle \citep{fruchter02},
          \textsc{eazy} \citep{brammer08},
          \textsc{GalfitM} \citep{peng10, hau2013},
          \textsc{Photutils} \citep{photutils},
          \textsc{Prospector} \citep{prospector},
          SciPy \citep{scipy},
          \textsc{Source Extractor} \citep{bertin96},
          \textsc{Statmorph} \citep{rod2019},
          STScI \textit{JWST} Calibration Pipeline (\url{jwst-pipeline.readthedocs.io})
}

\bibliographystyle{apj}

\begin{thebibliography}{101}
\expandafter\ifx\csname natexlab\endcsname\relax\def\natexlab#1{#1}\fi

\bibitem[{{Astropy Collaboration} {et~al.}(2013){Astropy Collaboration},
  {Robitaille}, {Tollerud}, {Greenfield}, {Droettboom}, {Bray}, {Aldcroft},
  {Davis}, {Ginsburg}, {Price-Whelan}, {Kerzendorf}, {Conley}, {Crighton},
  {Barbary}, {Muna}, {Ferguson}, {Grollier}, {Parikh}, {Nair}, {Unther},
  {Deil}, {Woillez}, {Conseil}, {Kramer}, {Turner}, {Singer}, {Fox}, {Weaver},
  {Zabalza}, {Edwards}, {Azalee Bostroem}, {Burke}, {Casey}, {Crawford},
  {Dencheva}, {Ely}, {Jenness}, {Labrie}, {Lim}, {Pierfederici}, {Pontzen},
  {Ptak}, {Refsdal}, {Servillat}, \& {Streicher}}]{astropy}
{Astropy Collaboration}, {Robitaille}, T.~P., {Tollerud}, E.~J., {et~al.} 2013,
  \aap, 558, A33

\bibitem[{{Bagley} {et~al.}(2022){Bagley}, {Finkelstein}, {Rojas-Ruiz},
  {Diekmann}, {Finkelstein}, {Song}, {Papovich}, {Somerville}, {Baronchelli},
  \& {Dai}}]{bagley22}
{Bagley}, M.~B., {Finkelstein}, S.~L., {Rojas-Ruiz}, S., {et~al.} 2022, arXiv
  e-prints, arXiv:2205.12980

\bibitem[{{Behroozi} {et~al.}(2020){Behroozi}, {Conroy}, {Wechsler}, {Hearin},
  {Williams}, {Moster}, {Yung}, {Somerville}, {Gottl{\"o}ber}, {Yepes}, \&
  {Endsley}}]{behroozi20}
{Behroozi}, P., {Conroy}, C., {Wechsler}, R.~H., {et~al.} 2020, \mnras, 499,
  5702

\bibitem[{{Behroozi} \& {Silk}(2015)}]{behroozi15}
{Behroozi}, P.~S., \& {Silk}, J. 2015, \apj, 799, 32

\bibitem[{{Bertin} \& {Arnouts}(1996)}]{bertin96}
{Bertin}, E., \& {Arnouts}, S. 1996, \aaps, 117, 393

\bibitem[{{Boquien} {et~al.}(2019){Boquien}, {Burgarella}, {Roehlly}, {Buat},
  {Ciesla}, {Corre}, {Inoue}, \& {Salas}}]{boquien19}
{Boquien}, M., {Burgarella}, D., {Roehlly}, Y., {et~al.} 2019, \aap, 622, A103

\bibitem[{{Bouwens} {et~al.}(2014){Bouwens}, {Illingworth}, {Oesch},
  {Labb{\'e}}, {van Dokkum}, {Trenti}, {Franx}, {Smit}, {Gonzalez}, \&
  {Magee}}]{bouwens14}
{Bouwens}, R.~J., {Illingworth}, G.~D., {Oesch}, P.~A., {et~al.} 2014, \apj,
  793, 115

\bibitem[{{Bouwens} {et~al.}(2015){Bouwens}, {Illingworth}, {Oesch}, {Trenti},
  {Labb{\'e}}, {Bradley}, {Carollo}, {van Dokkum}, {Gonzalez}, {Holwerda},
  {Franx}, {Spitler}, {Smit}, \& {Magee}}]{bouwens15}
---. 2015, \apj, 803, 34

\bibitem[{{Bouwens} {et~al.}(2021){Bouwens}, {Oesch}, {Stefanon},
  {Illingworth}, {Labb{\'e}}, {Reddy}, {Atek}, {Montes}, {Naidu},
  {Nanayakkara}, {Nelson}, \& {Wilkins}}]{bouwens21}
{Bouwens}, R.~J., {Oesch}, P.~A., {Stefanon}, M., {et~al.} 2021, \aj, 162, 47

\bibitem[{{Bowler} {et~al.}(2020){Bowler}, {Jarvis}, {Dunlop}, {McLure},
  {McLeod}, {Adams}, {Milvang-Jensen}, \& {McCracken}}]{bowler20}
{Bowler}, R.~A.~A., {Jarvis}, M.~J., {Dunlop}, J.~S., {et~al.} 2020, \mnras,
  493, 2059

\bibitem[{{Bowman} {et~al.}(2018){Bowman}, {Rogers}, {Monsalve}, {Mozdzen}, \&
  {Mahesh}}]{bowman18}
{Bowman}, J.~D., {Rogers}, A.~E.~E., {Monsalve}, R.~A., {Mozdzen}, T.~J., \&
  {Mahesh}, N. 2018, \nat, 555, 67

\bibitem[{{Bradley} {et~al.}(2020){Bradley}, {Sip{\H{o}}cz}, {Robitaille},
  {Tollerud}, {Vin{\'\i}cius}, {Deil}, {Barbary}, {Wilson}, {Busko},
  {G{\"u}nther}, {Cara}, {Conseil}, {Bostroem}, {Droettboom}, {Bray}, {Andersen
  Bratholm}, {Lim}, {Barentsen}, {Craig}, {Pascual}, {Perren}, {Greco},
  {Donath}, {de Val-Borro}, {Kerzendorf}, {Bach}, {Weaver}, {D'Eugenio},
  {Souchereau}, \& {Ferreira}}]{photutils}
{Bradley}, L., {Sip{\H{o}}cz}, B., {Robitaille}, T., {et~al.} 2020,
  {astropy/photutils: 1.0.0}, Zenodo

\bibitem[{{Brammer} {et~al.}(2008){Brammer}, {van Dokkum}, \&
  {Coppi}}]{brammer08}
{Brammer}, G.~B., {van Dokkum}, P.~G., \& {Coppi}, P. 2008, \apj, 686, 1503

\bibitem[{{Bruzual} \& {Charlot}(2003)}]{bruzual03}
{Bruzual}, G., \& {Charlot}, S. 2003, \mnras, 344, 1000

\bibitem[{{Burgarella} {et~al.}(2005){Burgarella}, {Buat}, \&
  {Iglesias-P{\'a}ramo}}]{burgarella05}
{Burgarella}, D., {Buat}, V., \& {Iglesias-P{\'a}ramo}, J. 2005, \mnras, 360,
  1413

\bibitem[{{Burgasser}(2014)}]{burgasser14}
{Burgasser}, A.~J. 2014, in Astronomical Society of India Conference Series,
  Vol.~11, Astronomical Society of India Conference Series, 7--16

\bibitem[{{Caballero} {et~al.}(2008){Caballero}, {Burgasser}, \&
  {Klement}}]{Caballero08}
{Caballero}, J.~A., {Burgasser}, A.~J., \& {Klement}, R. 2008, \aap, 488, 181

\bibitem[{{Caplar} \& {Tacchella}(2019)}]{caplar19}
{Caplar}, N., \& {Tacchella}, S. 2019, \mnras, 487, 3845

\bibitem[{{Cardelli} {et~al.}(1989){Cardelli}, {Clayton}, \&
  {Mathis}}]{cardelli89}
{Cardelli}, J.~A., {Clayton}, G.~C., \& {Mathis}, J.~S. 1989, \apj, 345, 245

\bibitem[{{Carnall} {et~al.}(2018){Carnall}, {McLure}, {Dunlop}, \&
  {Dav{\'e}}}]{carnall18}
{Carnall}, A.~C., {McLure}, R.~J., {Dunlop}, J.~S., \& {Dav{\'e}}, R. 2018,
  \mnras, 480, 4379

\bibitem[{{Casertano} {et~al.}(2000){Casertano}, {de Mello}, {Dickinson},
  {Ferguson}, {Fruchter}, {Gonzalez-Lopezlira}, {Heyer}, {Hook}, {Levay},
  {Lucas}, {Mack}, {Makidon}, {Mutchler}, {Smith}, {Stiavelli}, {Wiggs}, \&
  {Williams}}]{casertano00}
{Casertano}, S., {de Mello}, D., {Dickinson}, M., {et~al.} 2000, \aj, 120, 2747

\bibitem[{{Casey} {et~al.}(2014){Casey}, {Scoville}, {Sanders}, {Lee},
  {Cooray}, {Finkelstein}, {Capak}, {Conley}, {De Zotti}, {Farrah}, {Fu}, {Le
  Floc'h}, {Ilbert}, {Ivison}, \& {Takeuchi}}]{casey14b}
{Casey}, C.~M., {Scoville}, N.~Z., {Sanders}, D.~B., {et~al.} 2014, ArXiv
  e-prints

\bibitem[{{Castellano} {et~al.}(2022){Castellano}, {Fontana}, {Treu},
  {Santini}, {Merlin}, {Leethochawalit}, {Trenti}, {Mestric}, {Vanzella},
  {Bonchi}, {Belfiori}, {Nonino}, {Paris}, {Polenta}, {Roberts-Borsani},
  {Boyett}, {Calabro}, {Glazebrook}, {Grillo}, {Mascia}, {Mason}, {Mercurio},
  {Morishita}, {Nanayakkara}, {Pentericci}, {Rosati}, {Vulcani}, {Wang}, \&
  {Yang}}]{castellano22}
{Castellano}, M., {Fontana}, A., {Treu}, T., {et~al.} 2022, arXiv e-prints,
  arXiv:2207.09436

\bibitem[{{Chabrier}(2003)}]{chabrier03}
{Chabrier}, G. 2003, \pasp, 115, 763

\bibitem[{{Choi} {et~al.}(2017){Choi}, {Conroy}, \& {Byler}}]{choi17}
{Choi}, J., {Conroy}, C., \& {Byler}, N. 2017, \apj, 838, 159

\bibitem[{{Coe} {et~al.}(2013){Coe}, {Zitrin}, {Carrasco}, {Shu}, {Zheng},
  {Postman}, {Bradley}, {Koekemoer}, {Bouwens}, {Broadhurst}, {Monna}, {Host},
  {Moustakas}, {Ford}, {Moustakas}, {van der Wel}, {Donahue}, {Rodney},
  {Ben{\'{\i}}tez}, {Jouvel}, {Seitz}, {Kelson}, \& {Rosati}}]{coe13}
{Coe}, D., {Zitrin}, A., {Carrasco}, M., {et~al.} 2013, \apj, 762, 32

\bibitem[{{Conroy} \& {Gunn}(2010)}]{conroy10}
{Conroy}, C., \& {Gunn}, J.~E. 2010, {FSPS: Flexible Stellar Population
  Synthesis}

\bibitem[{{Dayal} \& {Ferrara}(2018)}]{dayal18}
{Dayal}, P., \& {Ferrara}, A. 2018, \physrep, 780, 1

\bibitem[{{Dayal} {et~al.}(2014){Dayal}, {Ferrara}, {Dunlop}, \&
  {Pacucci}}]{Dayal2014}
{Dayal}, P., {Ferrara}, A., {Dunlop}, J.~S., \& {Pacucci}, F. 2014, \mnras,
  445, 2545

\bibitem[{{Dayal} {et~al.}(2022){Dayal}, {Ferrara}, {Sommovigo}, {Bouwens},
  {Oesch}, {Smit}, {Gonzalez}, {Schouws}, {Stefanon}, {Kobayashi}, {Bremer},
  {Algera}, {Aravena}, {Bowler}, {da Cunha}, {Fudamoto}, {Graziani}, {Hodge},
  {Inami}, {De Looze}, {Pallottini}, {Riechers}, {Schneider}, {Stark}, \&
  {Endsley}}]{Dayal2022}
{Dayal}, P., {Ferrara}, A., {Sommovigo}, L., {et~al.} 2022, \mnras, 512, 989

\bibitem[{{Dunlop} {et~al.}(2013){Dunlop}, {Rogers}, {McLure}, {Ellis},
  {Robertson}, {Koekemoer}, {Dayal}, {Curtis-Lake}, {Wild}, {Charlot},
  {Bowler}, {Schenker}, {Ouchi}, {Ono}, {Cirasuolo}, {Furlanetto}, {Stark},
  {Targett}, \& {Schneider}}]{dunlop13}
{Dunlop}, J.~S., {Rogers}, A.~B., {McLure}, R.~J., {et~al.} 2013, \mnras, 432,
  3520

\bibitem[{{Eldridge} \& {Stanway}(2009)}]{eldridge09}
{Eldridge}, J.~J., \& {Stanway}, E.~R. 2009, \mnras, 400, 1019

\bibitem[{{Feng} {et~al.}(2016){Feng}, {Di-Matteo}, {Croft}, {Bird},
  {Battaglia}, \& {Wilkins}}]{Bluetides-I}
{Feng}, Y., {Di-Matteo}, T., {Croft}, R.~A., {et~al.} 2016, \mnras, 455, 2778

\bibitem[{{Ferland} {et~al.}(1998){Ferland}, {Korista}, {Verner}, {Ferguson},
  {Kingdon}, \& {Verner}}]{cloudy}
{Ferland}, G.~J., {Korista}, K.~T., {Verner}, D.~A., {et~al.} 1998, \pasp, 110,
  761

\bibitem[{{Finkelstein}(2016)}]{finkelstein16}
{Finkelstein}, S.~L. 2016, \pasa, 33, e037

\bibitem[{{Finkelstein} \& {Bagley}(2022)}]{finkelstein22b}
{Finkelstein}, S.~L., \& {Bagley}, M.~B. 2022, arXiv e-prints, arXiv:2207.02233

\bibitem[{{Finkelstein} {et~al.}(2012){Finkelstein}, {Papovich}, {Salmon},
  {Finlator}, {Dickinson}, {Ferguson}, {Giavalisco}, {Koekemoer}, {Reddy},
  {Bassett}, {Conselice}, {Dunlop}, {Faber}, {Grogin}, {Hathi}, {Kocevski},
  {Lai}, {Lee}, {McLure}, {Mobasher}, \& {Newman}}]{finkelstein12a}
{Finkelstein}, S.~L., {Papovich}, C., {Salmon}, B., {et~al.} 2012, \apj, 756,
  164

\bibitem[{{Finkelstein} {et~al.}(2015){Finkelstein}, {Ryan}, {Papovich},
  {Dickinson}, {Song}, {Somerville}, {Ferguson}, {Salmon}, {Giavalisco},
  {Koekemoer}, {Ashby}, {Behroozi}, {Castellano}, {Dunlop}, {Faber}, {Fazio},
  {Fontana}, {Grogin}, {Hathi}, {Jaacks}, {Kocevski}, {Livermore}, {McLure},
  {Merlin}, {Mobasher}, {Newman}, {Rafelski}, {Tilvi}, \&
  {Willner}}]{finkelstein15}
{Finkelstein}, S.~L., {Ryan}, Jr., R.~E., {Papovich}, C., {et~al.} 2015, \apj,
  810, 71

\bibitem[{{Finkelstein} {et~al.}(2022){Finkelstein}, {Bagley}, {Song},
  {Larson}, {Papovich}, {Dickinson}, {Finkelstein}, {Koekemoer}, {Pirzkal},
  {Somerville}, {Yung}, {Behroozi}, {Ferguson}, {Giavalisco}, {Grogin},
  {Hathi}, {Hutchison}, {Jung}, {Kocevski}, {Kawinwanichakij}, {Rojas-Ruiz},
  {Ryan}, {Snyder}, \& {Tacchella}}]{finkelstein22}
{Finkelstein}, S.~L., {Bagley}, M., {Song}, M., {et~al.} 2022, \apj, 928, 52

\bibitem[{{Flores Vel{\'a}zquez} {et~al.}(2021){Flores Vel{\'a}zquez},
  {Gurvich}, {Faucher-Gigu{\`e}re}, {Bullock}, {Starkenburg}, {Moreno},
  {Lazar}, {Mercado}, {Stern}, {Sparre}, {Hayward}, {Wetzel}, \&
  {El-Badry}}]{velazquez21}
{Flores Vel{\'a}zquez}, J.~A., {Gurvich}, A.~B., {Faucher-Gigu{\`e}re}, C.-A.,
  {et~al.} 2021, \mnras, 501, 4812

\bibitem[{{Fruchter} \& {Hook}(2002)}]{fruchter02}
{Fruchter}, A.~S., \& {Hook}, R.~N. 2002, \pasp, 114, 144

\bibitem[{{Geach} {et~al.}(2017){Geach}, {Dunlop}, {Halpern}, {Smail}, {van der
  Werf}, {Alexander}, {Almaini}, {Aretxaga}, {Arumugam}, {Asboth}, {Banerji},
  {Beanlands}, {Best}, {Blain}, {Birkinshaw}, {Chapin}, {Chapman}, {Chen},
  {Chrysostomou}, {Clarke}, {Clements}, {Conselice}, {Coppin}, {Cowley},
  {Danielson}, {Eales}, {Edge}, {Farrah}, {Gibb}, {Harrison}, {Hine}, {Hughes},
  {Ivison}, {Jarvis}, {Jenness}, {Jones}, {Karim}, {Koprowski}, {Knudsen},
  {Lacey}, {Mackenzie}, {Marsden}, {McAlpine}, {McMahon}, {Meijerink},
  {Micha{\l}owski}, {Oliver}, {Page}, {Peacock}, {Rigopoulou}, {Robson},
  {Roseboom}, {Rotermund}, {Scott}, {Serjeant}, {Simpson}, {Simpson}, {Smith},
  {Spaans}, {Stanley}, {Stevens}, {Swinbank}, {Targett}, {Thomson}, {Valiante},
  {Wake}, {Webb}, {Willott}, {Zavala}, \& {Zemcov}}]{geach17}
{Geach}, J.~E., {Dunlop}, J.~S., {Halpern}, M., {et~al.} 2017, \mnras, 465,
  1789

\bibitem[{{Gnedin}(2016)}]{gnedin16}
{Gnedin}, N.~Y. 2016, \apjl, 825, L17

\bibitem[{{Greene} {et~al.}(2016){Greene}, {Chu}, {Egami}, {Hodapp}, {Kelly},
  {Leisenring}, {Rieke}, {Robberto}, {Schlawin}, \& {Stansberry}}]{nircamgrism}
{Greene}, T.~P., {Chu}, L., {Egami}, E., {et~al.} 2016, in Society of
  Photo-Optical Instrumentation Engineers (SPIE) Conference Series, Vol. 9904,
  Space Telescopes and Instrumentation 2016: Optical, Infrared, and Millimeter
  Wave, ed. H.~A. {MacEwen}, G.~G. {Fazio}, M.~{Lystrup}, N.~{Batalha},
  N.~{Siegler}, \& E.~C. {Tong}, 99040E

\bibitem[{{Grogin} {et~al.}(2011){Grogin}, {Kocevski}, {Faber}, {Ferguson},
  {Koekemoer}, {Riess}, {Acquaviva}, {Alexander}, {Almaini}, {Ashby}, {Barden},
  {Bell}, {Bournaud}, {Brown}, {Caputi}, {Casertano}, {Cassata}, {Castellano},
  {Challis}, {Chary}, {Cheung}, {Cirasuolo}, {Conselice}, {Roshan Cooray},
  {Croton}, {Daddi}, {Dahlen}, {Dav{\'e}}, {de Mello}, {Dekel}, {Dickinson},
  {Dolch}, {Donley}, {Dunlop}, {Dutton}, {Elbaz}, {Fazio}, {Filippenko},
  {Finkelstein}, {Fontana}, {Gardner}, {Garnavich}, {Gawiser}, {Giavalisco},
  {Grazian}, {Guo}, {Hathi}, {H{\"a}ussler}, {Hopkins}, {Huang}, {Huang},
  {Jha}, {Kartaltepe}, {Kirshner}, {Koo}, {Lai}, {Lee}, {Li}, {Lotz}, {Lucas},
  {Madau}, {McCarthy}, {McGrath}, {McIntosh}, {McLure}, {Mobasher},
  {Moustakas}, {Mozena}, {Nandra}, {Newman}, {Niemi}, {Noeske}, {Papovich},
  {Pentericci}, {Pope}, {Primack}, {Rajan}, {Ravindranath}, {Reddy}, {Renzini},
  {Rix}, {Robaina}, {Rodney}, {Rosario}, {Rosati}, {Salimbeni}, {Scarlata},
  {Siana}, {Simard}, {Smidt}, {Somerville}, {Spinrad}, {Straughn}, {Strolger},
  {Telford}, {Teplitz}, {Trump}, {van der Wel}, {Villforth}, {Wechsler},
  {Weiner}, {Wiklind}, {Wild}, {Wilson}, {Wuyts}, {Yan}, \& {Yun}}]{grogin11}
{Grogin}, N.~A., {Kocevski}, D.~D., {Faber}, S.~M., {et~al.} 2011, \apjs, 197,
  35

\bibitem[{{Hainline} {et~al.}(2020){Hainline}, {Hviding}, {Rieke}, {Shivaei},
  {Endsley}, {Curtis-Lake}, {Smit}, {Williams}, {Alberts}, {K Boyett},
  {Bunker}, {Egami}, {Maseda}, {Tacchella}, \& {Willmer}}]{hainline20}
{Hainline}, K.~N., {Hviding}, R.~E., {Rieke}, M., {et~al.} 2020, \apj, 892, 125

\bibitem[{{Harikane} {et~al.}(2022){Harikane}, {Inoue}, {Mawatari},
  {Hashimoto}, {Yamanaka}, {Fudamoto}, {Matsuo}, {Tamura}, {Dayal}, {Yung},
  {Hutter}, {Pacucci}, {Sugahara}, \& {Koekemoer}}]{harikane22}
{Harikane}, Y., {Inoue}, A.~K., {Mawatari}, K., {et~al.} 2022, \apj, 929, 1

\bibitem[{{Hashimoto} {et~al.}(2018){Hashimoto}, {Laporte}, {Mawatari},
  {Ellis}, {Inoue}, {Zackrisson}, {Roberts-Borsani}, {Zheng}, {Tamura},
  {Bauer}, {Fletcher}, {Harikane}, {Hatsukade}, {Hayatsu}, {Matsuda}, {Matsuo},
  {Okamoto}, {Ouchi}, {Pell{\'o}}, {Rydberg}, {Shimizu}, {Taniguchi},
  {Umehata}, \& {Yoshida}}]{hashimoto18}
{Hashimoto}, T., {Laporte}, N., {Mawatari}, K., {et~al.} 2018, Nature, 557, 392

\bibitem[{{H{\"a}u{\ss}ler} {et~al.}(2013){H{\"a}u{\ss}ler}, {Bamford}, {Vika},
  {Rojas}, {Barden}, {Kelvin}, {Alpaslan}, {Robotham}, {Driver}, \&
  {Baldry}}]{hau2013}
{H{\"a}u{\ss}ler}, B., {Bamford}, S.~P., {Vika}, M., {et~al.} 2013, \mnras,
  430, 330

\bibitem[{{Iyer} \& {Gawiser}(2017)}]{iyer17}
{Iyer}, K., \& {Gawiser}, E. 2017, \apj, 838, 127

\bibitem[{{Iyer} {et~al.}(2019){Iyer}, {Gawiser}, {Faber}, {Ferguson},
  {Kartaltepe}, {Koekemoer}, {Pacifici}, \& {Somerville}}]{iyer19}
{Iyer}, K.~G., {Gawiser}, E., {Faber}, S.~M., {et~al.} 2019, \apj, 879, 116

\bibitem[{{Jakobsen} {et~al.}(2022){Jakobsen}, {Ferruit}, {Alves de Oliveira},
  {Arribas}, {Bagnasco}, {Barho}, {Beck}, {Birkmann}, {B{\"o}ker}, {Bunker},
  {Charlot}, {de Jong}, {de Marchi}, {Ehrenwinkler}, {Falcolini}, {Fels},
  {Franx}, {Franz}, {Funke}, {Giardino}, {Gnata}, {Holota}, {Honnen}, {Jensen},
  {Jentsch}, {Johnson}, {Jollet}, {Karl}, {Kling}, {K{\"o}hler}, {Kolm},
  {Kumari}, {Lander}, {Lemke}, {L{\'o}pez-Caniego}, {L{\"u}tzgendorf},
  {Maiolino}, {Manjavacas}, {Marston}, {Maschmann}, {Maurer}, {Messerschmidt},
  {Moseley}, {Mosner}, {Mott}, {Muzerolle}, {Pirzkal}, {Pittet}, {Plitzke},
  {Posselt}, {Rapp}, {Rauscher}, {Rawle}, {Rix}, {R{\"o}del}, {Rumler},
  {Sabbi}, {Salvignol}, {Schmid}, {Sirianni}, {Smith}, {Strada}, {te Plate},
  {Valenti}, {Wettemann}, {Wiehe}, {Wiesmayer}, {Willott}, {Wright}, {Zeidler},
  \& {Zincke}}]{nirspec}
{Jakobsen}, P., {Ferruit}, P., {Alves de Oliveira}, C., {et~al.} 2022, \aap,
  661, A80

\bibitem[{{Jiang} {et~al.}(2021){Jiang}, {Kashikawa}, {Wang}, {Walth}, {Ho},
  {Cai}, {Egami}, {Fan}, {Ito}, {Liang}, {Schaerer}, \& {Stark}}]{jiang21}
{Jiang}, L., {Kashikawa}, N., {Wang}, S., {et~al.} 2021, Nature Astronomy, 5,
  256

\bibitem[{{Johnson} {et~al.}(2021){Johnson}, {Leja}, {Conroy}, \&
  {Speagle}}]{prospector}
{Johnson}, B.~D., {Leja}, J., {Conroy}, C., \& {Speagle}, J.~S. 2021, \apjs,
  254, 22

\bibitem[{{Koekemoer} {et~al.}(2011){Koekemoer}, {Faber}, {Ferguson}, {Grogin},
  {Kocevski}, {Koo}, {Lai}, {Lotz}, {Lucas}, {McGrath}, {Ogaz}, {Rajan},
  {Riess}, {Rodney}, {Strolger}, {Casertano}, {Castellano}, {Dahlen},
  {Dickinson}, {Dolch}, {Fontana}, {Giavalisco}, {Grazian}, {Guo}, {Hathi},
  {Huang}, {van der Wel}, {Yan}, {Acquaviva}, {Alexander}, {Almaini}, {Ashby},
  {Barden}, {Bell}, {Bournaud}, {Brown}, {Caputi}, {Cassata}, {Challis},
  {Chary}, {Cheung}, {Cirasuolo}, {Conselice}, {Roshan Cooray}, {Croton},
  {Daddi}, {Dav{\'e}}, {de Mello}, {de Ravel}, {Dekel}, {Donley}, {Dunlop},
  {Dutton}, {Elbaz}, {Fazio}, {Filippenko}, {Finkelstein}, {Frazer}, {Gardner},
  {Garnavich}, {Gawiser}, {Gruetzbauch}, {Hartley}, {H{\"a}ussler},
  {Herrington}, {Hopkins}, {Huang}, {Jha}, {Johnson}, {Kartaltepe},
  {Khostovan}, {Kirshner}, {Lani}, {Lee}, {Li}, {Madau}, {McCarthy},
  {McIntosh}, {McLure}, {McPartland}, {Mobasher}, {Moreira}, {Mortlock},
  {Moustakas}, {Mozena}, {Nandra}, {Newman}, {Nielsen}, {Niemi}, {Noeske},
  {Papovich}, {Pentericci}, {Pope}, {Primack}, {Ravindranath}, {Reddy},
  {Renzini}, {Rix}, {Robaina}, {Rosario}, {Rosati}, {Salimbeni}, {Scarlata},
  {Siana}, {Simard}, {Smidt}, {Snyder}, {Somerville}, {Spinrad}, {Straughn},
  {Telford}, {Teplitz}, {Trump}, {Vargas}, {Villforth}, {Wagner}, {Wandro},
  {Wechsler}, {Weiner}, {Wiklind}, {Wild}, {Wilson}, {Wuyts}, \&
  {Yun}}]{koekemoer11}
{Koekemoer}, A.~M., {Faber}, S.~M., {Ferguson}, H.~C., {et~al.} 2011, \apjs,
  197, 36

\bibitem[{{Laporte} {et~al.}(2021){Laporte}, {Meyer}, {Ellis}, {Robertson},
  {Chisholm}, \& {Roberts-Borsani}}]{laporte21}
{Laporte}, N., {Meyer}, R.~A., {Ellis}, R.~S., {et~al.} 2021, \mnras, 505, 3336

\bibitem[{{Lovell} {et~al.}(2021){Lovell}, {Vijayan}, {Thomas}, {Wilkins},
  {Barnes}, {Irodotou}, \& {Roper}}]{FLARES-I}
{Lovell}, C.~C., {Vijayan}, A.~P., {Thomas}, P.~A., {et~al.} 2021, \mnras, 500,
  2127

\bibitem[{{Lutz} {et~al.}(2011){Lutz}, {Poglitsch}, {Altieri}, {Andreani},
  {Aussel}, {Berta}, {Bongiovanni}, {Brisbin}, {Cava}, {Cepa}, {Cimatti},
  {Daddi}, {Dominguez-Sanchez}, {Elbaz}, {F{\"o}rster Schreiber}, {Genzel},
  {Grazian}, {Gruppioni}, {Harwit}, {Le Floc'h}, {Magdis}, {Magnelli},
  {Maiolino}, {Nordon}, {P{\'e}rez Garc{\'\i}a}, {Popesso}, {Pozzi},
  {Riguccini}, {Rodighiero}, {Saintonge}, {Sanchez Portal}, {Santini}, {Shao},
  {Sturm}, {Tacconi}, {Valtchanov}, {Wetzstein}, \& {Wieprecht}}]{lutz11}
{Lutz}, D., {Poglitsch}, A., {Altieri}, B., {et~al.} 2011, \aap, 532, A90

\bibitem[{{Madau} \& {Dickinson}(2014)}]{madau14}
{Madau}, P., \& {Dickinson}, M. 2014, \araa, 52, 415

\bibitem[{{Magnelli} {et~al.}(2009){Magnelli}, {Elbaz}, {Chary}, {Dickinson},
  {Le Borgne}, {Frayer}, \& {Willmer}}]{magnelli09}
{Magnelli}, B., {Elbaz}, D., {Chary}, R.~R., {et~al.} 2009, \aap, 496, 57

\bibitem[{{Mason} {et~al.}(2015){Mason}, {Trenti}, \& {Treu}}]{Mason2015}
{Mason}, C.~A., {Trenti}, M., \& {Treu}, T. 2015, \apj, 813, 21

\bibitem[{{McLeod} {et~al.}(2016){McLeod}, {McLure}, \& {Dunlop}}]{mcleod16}
{McLeod}, D.~J., {McLure}, R.~J., \& {Dunlop}, J.~S. 2016, \mnras, 459, 3812

\bibitem[{{Morishita} {et~al.}(2018){Morishita}, {Trenti}, {Stiavelli},
  {Bradley}, {Coe}, {Oesch}, {Mason}, {Bridge}, {Holwerda}, {Livermore},
  {Salmon}, {Schmidt}, {Shull}, \& {Treu}}]{morishita18}
{Morishita}, T., {Trenti}, M., {Stiavelli}, M., {et~al.} 2018, \apj, 867, 150

\bibitem[{{Naidu} {et~al.}(2022){Naidu}, {Oesch}, {van Dokkum}, {Nelson},
  {Suess}, {Whitaker}, {Allen}, {Bezanson}, {Bouwens}, {Brammer}, {Conroy},
  {Illingworth}, {Labbe}, {Leja}, {Leonova}, {Matthee}, {Price}, {Setton},
  {Strait}, {Stefanon}, {Tacchella}, {Toft}, {Weaver}, \& {Weibel}}]{naidu22}
{Naidu}, R.~P., {Oesch}, P.~A., {van Dokkum}, P., {et~al.} 2022, arXiv
  e-prints, arXiv:2207.09434

\bibitem[{{Noll} {et~al.}(2009){Noll}, {Burgarella}, {Giovannoli}, {Buat},
  {Marcillac}, \& {Mu{\~n}oz-Mateos}}]{noll09}
{Noll}, S., {Burgarella}, D., {Giovannoli}, E., {et~al.} 2009, \aap, 507, 1793

\bibitem[{{Oesch} {et~al.}(2018){Oesch}, {Bouwens}, {Illingworth}, {Labb{\'e}},
  \& {Stefanon}}]{oesch18}
{Oesch}, P.~A., {Bouwens}, R.~J., {Illingworth}, G.~D., {Labb{\'e}}, I., \&
  {Stefanon}, M. 2018, \apj, 855, 105

\bibitem[{{Oesch} {et~al.}(2016){Oesch}, {Brammer}, {van Dokkum},
  {Illingworth}, {Bouwens}, {Labb{\'e}}, {Franx}, {Momcheva}, {Ashby}, {Fazio},
  {Gonzalez}, {Holden}, {Magee}, {Skelton}, {Smit}, {Spitler}, {Trenti}, \&
  {Willner}}]{oesch16}
{Oesch}, P.~A., {Brammer}, G., {van Dokkum}, P.~G., {et~al.} 2016, \apj, 819,
  129

\bibitem[{{Oke} \& {Gunn}(1983)}]{oke83}
{Oke}, J.~B., \& {Gunn}, J.~E. 1983, \apj, 266, 713

\bibitem[{{Oliver} {et~al.}(2012){Oliver}, {Bock}, {Altieri}, {Amblard},
  {Arumugam}, {Aussel}, {Babbedge}, {Beelen}, {B{\'e}thermin}, {Blain},
  {Boselli}, {Bridge}, {Brisbin}, {Buat}, {Burgarella},
  {Castro-Rodr{\'\i}guez}, {Cava}, {Chanial}, {Cirasuolo}, {Clements},
  {Conley}, {Conversi}, {Cooray}, {Dowell}, {Dubois}, {Dwek}, {Dye}, {Eales},
  {Elbaz}, {Farrah}, {Feltre}, {Ferrero}, {Fiolet}, {Fox}, {Franceschini},
  {Gear}, {Giovannoli}, {Glenn}, {Gong}, {Gonz{\'a}lez Solares}, {Griffin},
  {Halpern}, {Harwit}, {Hatziminaoglou}, {Heinis}, {Hurley}, {Hwang}, {Hyde},
  {Ibar}, {Ilbert}, {Isaak}, {Ivison}, {Lagache}, {Le Floc'h}, {Levenson},
  {Faro}, {Lu}, {Madden}, {Maffei}, {Magdis}, {Mainetti}, {Marchetti},
  {Marsden}, {Marshall}, {Mortier}, {Nguyen}, {O'Halloran}, {Omont}, {Page},
  {Panuzzo}, {Papageorgiou}, {Patel}, {Pearson}, {P{\'e}rez-Fournon}, {Pohlen},
  {Rawlings}, {Raymond}, {Rigopoulou}, {Riguccini}, {Rizzo}, {Rodighiero},
  {Roseboom}, {Rowan-Robinson}, {S{\'a}nchez Portal}, {Schulz}, {Scott},
  {Seymour}, {Shupe}, {Smith}, {Stevens}, {Symeonidis}, {Trichas}, {Tugwell},
  {Vaccari}, {Valtchanov}, {Vieira}, {Viero}, {Vigroux}, {Wang}, {Ward},
  {Wardlow}, {Wright}, {Xu}, \& {Zemcov}}]{oliver12}
{Oliver}, S.~J., {Bock}, J., {Altieri}, B., {et~al.} 2012, \mnras, 424, 1614

\bibitem[{{Papovich} {et~al.}(2016){Papovich}, {Shipley}, {Mehrtens}, {Lanham},
  {Lacy}, {Ciardullo}, {Finkelstein}, {Bassett}, {Behroozi}, {Blanc}, {de
  Jong}, {DePoy}, {Drory}, {Gawiser}, {Gebhardt}, {Gronwall}, {Hill}, {Hopp},
  {Jogee}, {Kawinwanichakij}, {Marshall}, {McLinden}, {Mentuch Cooper},
  {Somerville}, {Steinmetz}, {Tran}, {Tuttle}, {Viero}, {Wechsler}, \&
  {Zeimann}}]{papovich16}
{Papovich}, C., {Shipley}, H.~V., {Mehrtens}, N., {et~al.} 2016, \apjs, 224, 28

\bibitem[{{Patten} {et~al.}(2006){Patten}, {Stauffer}, {Burrows}, {Marengo},
  {Hora}, {Luhman}, {Sonnett}, {Henry}, {Raghavan}, \& {Megeath}}]{patten06}
{Patten}, B.~M., {Stauffer}, J.~R., {Burrows}, A., {et~al.} 2006, \apj, 651,
  502

\bibitem[{{Peng} {et~al.}(2002){Peng}, {Ho}, {Impey}, \& {Rix}}]{peng2002}
{Peng}, C.~Y., {Ho}, L.~C., {Impey}, C.~D., \& {Rix}, H.-W. 2002, \aj, 124, 266

\bibitem[{Peng {et~al.}(2010)Peng, {Ho}, {Impey}, \& {Rix}}]{peng2010}
Peng, C.~Y., {Ho}, L.~C., {Impey}, C.~D., \& {Rix}, H.-W. 2010, \aj, 139, 2097

\bibitem[{{Peng} {et~al.}(2010){Peng}, {Ho}, {Impey}, \& {Rix}}]{peng10}
{Peng}, C.~Y., {Ho}, L.~C., {Impey}, C.~D., \& {Rix}, H.-W. 2010, \aj, 139,
  2097

\bibitem[{{P{\'e}rez-Gonz{\'a}lez} {et~al.}(2008){P{\'e}rez-Gonz{\'a}lez},
  {Rieke}, {Villar}, {Barro}, {Blaylock}, {Egami}, {Gallego}, {Gil de Paz},
  {Pascual}, {Zamorano}, \& {Donley}}]{perezgonzalez08}
{P{\'e}rez-Gonz{\'a}lez}, P.~G., {Rieke}, G.~H., {Villar}, V., {et~al.} 2008,
  \apj, 675, 234

\bibitem[{{Planck Collaboration} {et~al.}(2020){Planck Collaboration},
  {Aghanim}, {Akrami}, {Ashdown}, {Aumont}, {Baccigalupi}, {Ballardini},
  {Banday}, {Barreiro}, {Bartolo}, {Basak}, {Battye}, {Benabed}, {Bernard},
  {Bersanelli}, {Bielewicz}, {Bock}, {Bond}, {Borrill}, {Bouchet}, {Boulanger},
  {Bucher}, {Burigana}, {Butler}, {Calabrese}, {Cardoso}, {Carron},
  {Challinor}, {Chiang}, {Chluba}, {Colombo}, {Combet}, {Contreras}, {Crill},
  {Cuttaia}, {de Bernardis}, {de Zotti}, {Delabrouille}, {Delouis}, {Di
  Valentino}, {Diego}, {Dor{\'e}}, {Douspis}, {Ducout}, {Dupac}, {Dusini},
  {Efstathiou}, {Elsner}, {En{\ss}lin}, {Eriksen}, {Fantaye}, {Farhang},
  {Fergusson}, {Fernandez-Cobos}, {Finelli}, {Forastieri}, {Frailis},
  {Fraisse}, {Franceschi}, {Frolov}, {Galeotta}, {Galli}, {Ganga},
  {G{\'e}nova-Santos}, {Gerbino}, {Ghosh}, {Gonz{\'a}lez-Nuevo}, {G{\'o}rski},
  {Gratton}, {Gruppuso}, {Gudmundsson}, {Hamann}, {Handley}, {Hansen},
  {Herranz}, {Hildebrandt}, {Hivon}, {Huang}, {Jaffe}, {Jones}, {Karakci},
  {Keih{\"a}nen}, {Keskitalo}, {Kiiveri}, {Kim}, {Kisner}, {Knox},
  {Krachmalnicoff}, {Kunz}, {Kurki-Suonio}, {Lagache}, {Lamarre}, {Lasenby},
  {Lattanzi}, {Lawrence}, {Le Jeune}, {Lemos}, {Lesgourgues}, {Levrier},
  {Lewis}, {Liguori}, {Lilje}, {Lilley}, {Lindholm}, {L{\'o}pez-Caniego},
  {Lubin}, {Ma}, {Mac{\'\i}as-P{\'e}rez}, {Maggio}, {Maino}, {Mandolesi},
  {Mangilli}, {Marcos-Caballero}, {Maris}, {Martin}, {Martinelli},
  {Mart{\'\i}nez-Gonz{\'a}lez}, {Matarrese}, {Mauri}, {McEwen}, {Meinhold},
  {Melchiorri}, {Mennella}, {Migliaccio}, {Millea}, {Mitra},
  {Miville-Desch{\^e}nes}, {Molinari}, {Montier}, {Morgante}, {Moss}, {Natoli},
  {N{\o}rgaard-Nielsen}, {Pagano}, {Paoletti}, {Partridge}, {Patanchon},
  {Peiris}, {Perrotta}, {Pettorino}, {Piacentini}, {Polastri}, {Polenta},
  {Puget}, {Rachen}, {Reinecke}, {Remazeilles}, {Renzi}, {Rocha}, {Rosset},
  {Roudier}, {Rubi{\~n}o-Mart{\'\i}n}, {Ruiz-Granados}, {Salvati}, {Sandri},
  {Savelainen}, {Scott}, {Shellard}, {Sirignano}, {Sirri}, {Spencer},
  {Sunyaev}, {Suur-Uski}, {Tauber}, {Tavagnacco}, {Tenti}, {Toffolatti},
  {Tomasi}, {Trombetti}, {Valenziano}, {Valiviita}, {Van Tent}, {Vibert},
  {Vielva}, {Villa}, {Vittorio}, {Wandelt}, {Wehus}, {White}, {White},
  {Zacchei}, \& {Zonca}}]{planck20}
{Planck Collaboration}, {Aghanim}, N., {Akrami}, Y., {et~al.} 2020, \aap, 641,
  A6

\bibitem[{{Rieke} {et~al.}(2015){Rieke}, {Wright}, {B{\"o}ker}, {Bouwman},
  {Colina}, {Glasse}, {Gordon}, {Greene}, {G{\"u}del}, {Henning}, {Justtanont},
  {Lagage}, {Meixner}, {N{\o}rgaard-Nielsen}, {Ray}, {Ressler}, {van Dishoeck},
  \& {Waelkens}}]{miri}
{Rieke}, G.~H., {Wright}, G.~S., {B{\"o}ker}, T., {et~al.} 2015, \pasp, 127,
  584

\bibitem[{{Rieke} {et~al.}(2005){Rieke}, {Kelly}, \& {Horner}}]{rieke05}
{Rieke}, M.~J., {Kelly}, D., \& {Horner}, S. 2005, in Society of Photo-Optical
  Instrumentation Engineers (SPIE) Conference Series, Vol. 5904, Cryogenic
  Optical Systems and Instruments XI, ed. J.~B. {Heaney} \& L.~G. {Burriesci},
  1--8

\bibitem[{{Robertson}(2021)}]{robertson21}
{Robertson}, B.~E. 2021, arXiv e-prints, arXiv:2110.13160

\bibitem[{{Rodriguez-Gomez} {et~al.}(2019){Rodriguez-Gomez}, {Snyder}, {Lotz},
  {Nelson}, {Pillepich}, {Springel}, {Genel}, {Weinberger}, {Tacchella}, \&
  {Pakmor}}]{rod2019}
{Rodriguez-Gomez}, V., {Snyder}, G.~F., {Lotz}, J.~M., {et~al.} 2019, \mnras,
  483, 4140

\bibitem[{{Rojas-Ruiz} {et~al.}(2020){Rojas-Ruiz}, {Finkelstein}, {Bagley},
  {Stevans}, {Finkelstein}, {Larson}, {Mechtley}, \& {Diekmann}}]{rojasruiz20}
{Rojas-Ruiz}, S., {Finkelstein}, S.~L., {Bagley}, M.~B., {et~al.} 2020, \apj,
  891, 146

\bibitem[{{Ryan} {et~al.}(2005){Ryan}, {Hathi}, {Cohen}, \&
  {Windhorst}}]{Ryan05}
{Ryan}, R.~E., J., {Hathi}, N.~P., {Cohen}, S.~H., \& {Windhorst}, R.~A. 2005,
  \apjl, 631, L159

\bibitem[{{Schlawin} {et~al.}(2020){Schlawin}, {Leisenring}, {Misselt},
  {Greene}, {McElwain}, {Beatty}, \& {Rieke}}]{schlawin20}
{Schlawin}, E., {Leisenring}, J., {Misselt}, K., {et~al.} 2020, \aj, 160, 231

\bibitem[{{Skelton} {et~al.}(2014){Skelton}, {Whitaker}, {Momcheva}, {Brammer},
  {van Dokkum}, {Labb{\'e}}, {Franx}, {van der Wel}, {Bezanson}, {Da Cunha},
  {Fumagalli}, {F{\"o}rster Schreiber}, {Kriek}, {Leja}, {Lundgren}, {Magee},
  {Marchesini}, {Maseda}, {Nelson}, {Oesch}, {Pacifici}, {Patel}, {Price},
  {Rix}, {Tal}, {Wake}, \& {Wuyts}}]{skelton14}
{Skelton}, R.~E., {Whitaker}, K.~E., {Momcheva}, I.~G., {et~al.} 2014, \apjs,
  214, 24

\bibitem[{{Somerville} {et~al.}(2021){Somerville}, {Olsen}, {Yung}, {Pacifici},
  {Ferguson}, {Behroozi}, {Osborne}, {Wechsler}, {Pandya}, {Faber}, {Primack},
  \& {Dekel}}]{somerville21}
{Somerville}, R.~S., {Olsen}, C., {Yung}, L.~Y.~A., {et~al.} 2021, \mnras, 502,
  4858

\bibitem[{{Stark}(2016)}]{stark16}
{Stark}, D.~P. 2016, \araa, 54, 761

\bibitem[{{Stefanon} {et~al.}(2022){Stefanon}, {Bouwens}, {Illingworth},
  {Labb{\'e}}, {Oesch}, \& {Gonzalez}}]{stefanon22}
{Stefanon}, M., {Bouwens}, R.~J., {Illingworth}, G.~D., {et~al.} 2022, arXiv
  e-prints, arXiv:2204.02986

\bibitem[{{Stefanon} {et~al.}(2017){Stefanon}, {Yan}, {Mobasher}, {Barro},
  {Donley}, {Fontana}, {Hemmati}, {Koekemoer}, {Lee}, {Lee}, {Nayyeri}, {Peth},
  {Pforr}, {Salvato}, {Wiklind}, {Wuyts}, {Ashby}, {Castellano}, {Conselice},
  {Cooper}, {Cooray}, {Dolch}, {Ferguson}, {Galametz}, {Giavalisco}, {Guo},
  {Willner}, {Dickinson}, {Faber}, {Fazio}, {Gardner}, {Gawiser}, {Grazian},
  {Grogin}, {Kocevski}, {Koo}, {Lee}, {Lucas}, {McGrath}, {Nandra}, {Newman},
  \& {van der Wel}}]{stefanon17}
{Stefanon}, M., {Yan}, H., {Mobasher}, B., {et~al.} 2017, \apjs, 229, 32

\bibitem[{{Sullivan} {et~al.}(2018){Sullivan}, {Hirano}, \&
  {Bromm}}]{sullivan18}
{Sullivan}, J.~M., {Hirano}, S., \& {Bromm}, V. 2018, \mnras, 481, L69

\bibitem[{{Tacchella} {et~al.}(2018){Tacchella}, {Bose}, {Conroy},
  {Eisenstein}, \& {Johnson}}]{tacchella18}
{Tacchella}, S., {Bose}, S., {Conroy}, C., {Eisenstein}, D.~J., \& {Johnson},
  B.~D. 2018, \apj, 868, 92

\bibitem[{{Tacchella} {et~al.}(2022){Tacchella}, {Finkelstein}, {Bagley},
  {Dickinson}, {Ferguson}, {Giavalisco}, {Graziani}, {Grogin}, {Hathi},
  {Hutchison}, {Jung}, {Koekemoer}, {Larson}, {Papovich}, {Pirzkal},
  {Rojas-Ruiz}, {Song}, {Schneider}, {Somerville}, {Wilkins}, \&
  {Yung}}]{tacchella22}
{Tacchella}, S., {Finkelstein}, S.~L., {Bagley}, M., {et~al.} 2022, \apj, 927,
  170

\bibitem[{{Vijayan} {et~al.}(2021){Vijayan}, {Lovell}, {Wilkins}, {Thomas},
  {Barnes}, {Irodotou}, {Kuusisto}, \& {Roper}}]{FLARES-II}
{Vijayan}, A.~P., {Lovell}, C.~C., {Wilkins}, S.~M., {et~al.} 2021, \mnras,
  501, 3289

\bibitem[{Virtanen {et~al.}(2020)Virtanen, Gommers, Oliphant, Haberland, Reddy,
  Cournapeau, Burovski, Peterson, Weckesser, Bright, {van der Walt}, Brett,
  Wilson, Millman, Mayorov, Nelson, Jones, Kern, Larson, Carey, Polat, Feng,
  Moore, {VanderPlas}, Laxalde, Perktold, Cimrman, Henriksen, Quintero, Harris,
  Archibald, Ribeiro, Pedregosa, {van Mulbregt}, \& {SciPy 1.0
  Contributors}}]{scipy}
Virtanen, P., Gommers, R., Oliphant, T.~E., {et~al.} 2020, Nature Methods, 17,
  261

\bibitem[{{Wilkins} {et~al.}(2017){Wilkins}, {Feng}, {Di Matteo}, {Croft},
  {Lovell}, \& {Waters}}]{Bluetides-II}
{Wilkins}, S.~M., {Feng}, Y., {Di Matteo}, T., {et~al.} 2017, \mnras, 469, 2517

\bibitem[{{Wilkins} {et~al.}(2014){Wilkins}, {Stanway}, \&
  {Bremer}}]{Wilkins2014}
{Wilkins}, S.~M., {Stanway}, E.~R., \& {Bremer}, M.~N. 2014, \mnras, 439, 1038

\bibitem[{{Wilkins} {et~al.}(2022){Wilkins}, {Vijayan}, {Lovell}, {Roper},
  {Irodotou}, {Caruana}, {Seeyave}, {Kuusisto}, {Thomas}, \&
  {Parris}}]{FLARES-V}
{Wilkins}, S.~M., {Vijayan}, A.~P., {Lovell}, C.~C., {et~al.} 2022, arXiv
  e-prints, arXiv:2204.09431

\bibitem[{{Yan} {et~al.}(2003){Yan}, {Windhorst}, \& {Cohen}}]{Yan03}
{Yan}, H., {Windhorst}, R.~A., \& {Cohen}, S.~H. 2003, \apjl, 585, L93

\bibitem[{{Yung} {et~al.}(2019){Yung}, {Somerville}, {Finkelstein}, {Popping},
  \& {Dav{\'e}}}]{yung19a}
{Yung}, L.~Y.~A., {Somerville}, R.~S., {Finkelstein}, S.~L., {Popping}, G., \&
  {Dav{\'e}}, R. 2019, \mnras, 483, 2983

\bibitem[{{Yung} {et~al.}(2020){Yung}, {Somerville}, {Finkelstein}, {Popping},
  {Dav{\'e}}, {Venkatesan}, {Behroozi}, \& {Ferguson}}]{yung20b}
{Yung}, L.~Y.~A., {Somerville}, R.~S., {Finkelstein}, S.~L., {et~al.} 2020,
  \mnras, 496, 4574

\bibitem[{{Yung} {et~al.}(2022){Yung}, {Somerville}, {Ferguson}, {Finkelstein},
  {Gardner}, {Dav{\'e}}, {Bagley}, {Popping}, \& {Behroozi}}]{yung22}
{Yung}, L.~Y.~A., {Somerville}, R.~S., {Ferguson}, H.~C., {et~al.} 2022, arXiv
  e-prints, arXiv:2206.13521

\bibitem[{{Zavala} {et~al.}(2022){Zavala}, {Buat}, {Casey}, {Burgarella},
  {Finkelstein}, {Bagley}, {Ciesla}, {Daddi}, {Dickinson}, {Ferguson},
  {Franco}, {Jim'enez-Andrade}, {Kartaltepe}, {Koekemoer}, {Le Bail}, {Murphy},
  {Papovich}, {Tacchella}, {Wilkins}, {Fontana}, {Giavalisco}, {Grazian},
  {Grogin}, {Kewley}, {Kocevski}, {Kirkpatrick}, {Lotz}, {Pentericci},
  {Perez-Gonzalez}, {Pirzkal}, {Ravindranath}, {Somerville}, {Trump}, {Yang},
  {Yung}, {Almaini}, {Amorin}, {Annunziatella}, {Arrabal Haro}, {Backhaus},
  {Barro}, {Behroozi}, {Bell}, {Bhatawdekar}, {Bisigello}, {Buitrago},
  {Calabro}, {Castellano}, {Chavez Ortiz}, {Chworowsky}, {Cleri}, {Cohen},
  {Cole}, {Cooke}, {Cooper}, {Cooray}, {Costantin}, {Cox}, {Croton}, {Dave},
  {de la Vega}, {Dekel}, {Elbaz}, {Estrada-Carpenter}, {Fern{\'a}ndez},
  {Finkelstein}, {Freundlich}, {Fujimoto}, {Garc{\'\i}a-Argum{\'a}nez},
  {Gardner}, {Gawiser}, {G{\'o}mez-Guijarro}, {Guo}, {Hamilton}, {Hathi},
  {Holwerda}, {Hirschmann}, {Huertas-Company}, {Hutchison}, {Iyer}, {Jaskot},
  {Jha}, {Jogee}, {Juneau}, {Jung}, {Kassin}, {Kurczynski}, {Larson}, {Leung},
  {Lucas}, {Magnelli}, {Mantha}, {Matharu}, {McGrath}, {McIntosh}, {Medrano},
  {Merlin}, {Mobasher}, {Morales}, {Newman}, {Nicholls}, {Pandya}, {Rafelski},
  {Ronayne}, {Rose}, {Ryan}, {Santini}, {Seill{\'e}}, {Shah}, {Shen}, {Simons},
  {Snyder}, {Stanway}, {Straughn}, {Teplitz}, {Vanderhoof}, {Vega-Ferrero},
  {Wang}, {Weiner}, {Willmer}, \& {Wuyts}}]{zavala22}
{Zavala}, J.~A., {Buat}, V., {Casey}, C.~M., {et~al.} 2022, arXiv e-prints,
  arXiv:2208.01816

\end{thebibliography}

\end{document}